\documentclass[usenatbib]{mn2e}

\usepackage{lscape}
\usepackage{graphicx}
\usepackage{amssymb}
\usepackage{url}
\usepackage{aas_macros}

\newcommand{\msun}{\mbox{$M_{\odot}$}}		

\usepackage{multirow}

\title[The Pleiades benchmark]{
  Pre-main-sequence isochrones -- I. The Pleiades benchmark}
\author[Cameron~P.~M.~Bell et al.]{Cameron~P.~M.~Bell$^{1}$\thanks{E-mail:
  bell@astro.ex.ac.uk}, Tim Naylor$^{1}$,
N.~J.~Mayne$^{1}$, R.~D.~Jeffries$^{2}$ and S. P. Littlefair$^{3}$\\
$^{1}$ School of Physics, University of Exeter, Exeter EX4 4QL\\
$^{2}$ Astrophysics Group, Research Institute for the Environment,
Physical Sciences and Applied Mathematics, Keele University,\\
Staffordshire ST5 5BG\\
$^{3}$ Department of Physics and Astronomy, University of Sheffield,
Sheffield S3 7RH}

\begin{document}

\date{Accepted ?, Received ?; in original form ?}

\pagerange{\pageref{firstpage}--\pageref{lastpage}} \pubyear{2012}

\maketitle

\label{firstpage}

\begin{abstract}
  We present a critical assessment of commonly used pre-main-sequence
  isochrones by comparing their predictions to a set of well-calibrated colour-magnitude
  diagrams of the Pleiades in the wavelength range 0.4 to $2.5\, \mu$m. 
  Our analysis shows that for temperatures less than $4000\, $K the models systematically overestimate the flux by a factor
  two at $0.5\, \mu$m, though this decreases with wavelength, becoming
  negligible at $2.2\, \mu$m.
  In optical colours this will result in the ages for stars younger than $10\, \rm{Myr}$ being
  underestimated by factors between two and three.

  We show that using observations of 
  standard stars to transform the data into a standard system can introduce significant 
  errors in the positioning of pre-main-sequences in colour-magnitude diagrams.
  Therefore we have compared the models to the data in the natural photometric system in which the
  observations were taken. 
  Thus we have constructed and tested a model of the system responses for the
  Wide-Field Camera on the \textit{Isaac Newton} Telescope.

  As a benchmark test for the development of pre-main-sequence models we provide both our
  system responses and the Pleiades sequence.
\end{abstract}


\section{Introduction}
\label{introduction}

Our understanding of time-dependent physical processes in young
pre-main-sequence (pre-MS) and main-sequence (MS)
populations is fundamentally limited by poorly constrained
timescales, the derivation of which requires the determination of
ages for pre-MS stellar populations.
These ages are generally derived via the comparison of photometric observations 
of pre-MS stars in young open clusters with theoretical models (or isochrones) using
colour-magnitude diagrams (CMDs).

Many previous studies deriving stellar parameters from observations of
young clusters in this way have shown large discrepancies between the
photometric data and the model predictions. 
It has been well documented that pre-MS isochrones generally do not fit the entire 
dataset in CMDs; e.g. whilst tracing the sequence defined by the higher mass members 
they will not simultaneously follow the sequence in the lower mass regime (see for 
instance \citealp{Hartmann03,Stauffer07}). 
In addition, pre-MS isochrones from different evolutionary models give ages that differ by a factor two for the same 
cluster (e.g. \citealp{Dahm05,Mayne07}).
Furthermore, the same dataset compared to isochrones in 
different colours can lead to different age determinations (e.g. \citealp{Naylor02}).

The situation has become even more uncertain with the realisation that ages
derived from the MS are systematically older by a factor 1.5 to 2 than ages derived
from the pre-MS (e.g. \citealp{Naylor09}).
To make further progress we require a precise test of the pre-MS models to determine whether the ages could be
in error by such a large factor, and if any parts of the pre-MS or any colour combinations are
reliable age indicators.
Our longer term aim is that if we have a clear understanding of the limitations of pre-MS ages we may be able to
create a reliable pre-MS age scale. 

In this work, therefore, we set up a benchmark test for pre-MS
isochrones. Most obviously that benchmark must be a cluster that
contains a significant number of pre-MS objects, as well as a
populated MS to show that the models at least fit these more evolved
stars.
In addition it should have distance and age determinations which are independent of the CMD,
and preferably a low extinction (which might otherwise complicate transformations).
Given that most pre-MS isochrones are only available for solar metallicity, the cluster must
be at least close to solar composition.
Finally, it should be nearby so that the lowest possible masses can be accessed.
The Pleiades is the only cluster which fits all these criteria, and \cite{Stauffer07} have already
undertaken a comparison in $V$ and $I_{\rm{c}}$, which shows that at least the $V$-band luminosity
is not matched by two models available at that time.

In this paper we return to the Pleiades with the aim of assessing which of the most recent models 
best match the data. This then allows us to quantify the remaining
mismatch so that we can make a realistic assessment of its impact on
pre-MS ages.
We trace the mismatch as a function of wavelength using observations in a 
set of filters which are contiguous in wavelength and cover 0.4 to
$2.5\, \mu$m.
The data reach cooler temperatures than the \cite{Stauffer07} study,
and are free of the uncertainties introduced by transforming heterogeneous
datasets into a standard photometric system.

In Section~\ref{the_models} we describe the set of interior and
atmospheric theoretical models we have
studied and examine systematic differences between the different
formalisms. In Section~\ref{the_data} we describe a  set of
observations, of both standard stars and the Pleiades, which we use to
test the models. In Section~\ref{testing_the_models_using_cmds} we
compare the models to the Pleiades dataset using
CMDs. Section~\ref{optical_nir_mass-luminosity_relations} describes the
comparison of the models to a sample of MS binaries with dynamically
constrained masses. In
Section~\ref{empirical_bolometric_corrections} we quantify the
mismatch between the models and the data as a function of
wavelength. Our conclusions are detailed in
Section~\ref{conclusions}. 

\section{The Models}
\label{the_models}

To derive fundamental
parameters, such as age and distance, from fitting CMDs we require
model isochrones, which must also be transformed into the requisite
photometric system. Isochrones are constructed from stellar interior models
that predict $L_{\rm{bol}}$, $T_{\rm{eff}}$ and surface gravity (log$\,g$).
Colours are then calculated through a
colour-$T_{\rm{eff}}$ relation and magnitudes via bolometric
corrections to $L_{\rm{bol}}$. Both relations can be
derived by folding atmospheric model flux distributions
through the appropriate
photometric filter responses. The colours and magnitudes must then be
calibrated to a standard scale. Therefore, for this study we have
adopted a set of pre-MS interior models and also a collection of
atmospheric models with which to derive the required relations.

\subsection{Pre-main-sequence interior models}
\label{pre-ms_models}

We tested those pre-MS interior models
that were publicly available, covered a wide range in stellar mass
and had been designed specifically for the investigation of pre-MS
evolution. The sets of interior models studied here are those of
\cite{Baraffe98}, \cite*{Siess00}, \cite{D'Antona97} and
\cite{Dotter08} (hereafter
BCAH98, SDF00, DAM97 and DCJ08 respectively\footnote[1]{Whilst the
  DCJ08 models are not specifically designed for
  studying pre-MS evolution, these mass tracks cover the entire pre-MS
  phase over a significant mass range.}). Note that the recent
Pisa models \citep*{Tognelli11} do not extend to the age of the
Pleiades (see Section~\ref{model_parameters}) and therefore are not
used in this study.

\begin{figure}
\centering
\includegraphics[width=\columnwidth]{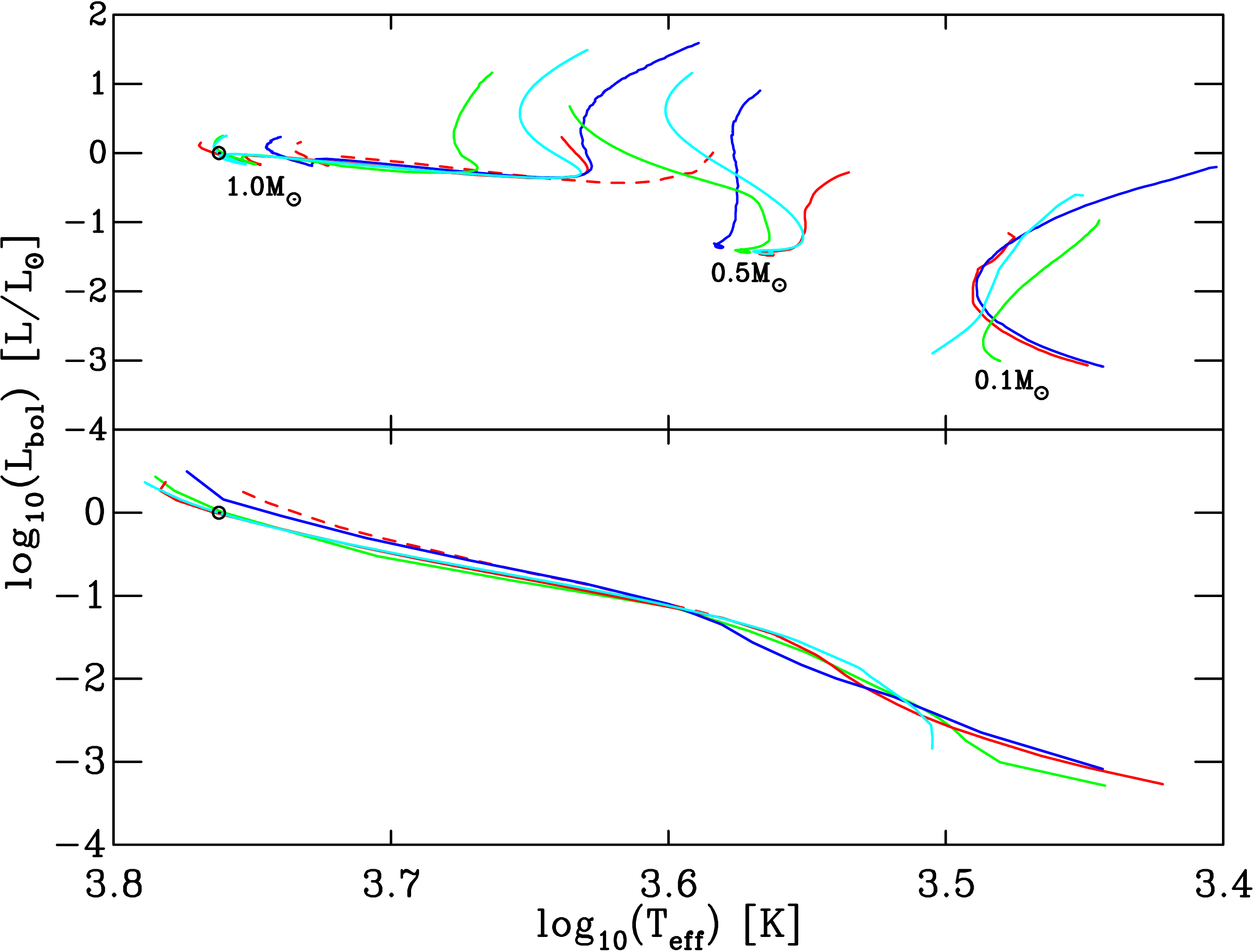}
\caption{Variation between the pre-MS evolutionary tracks for
  masses of $0.1, 0.5$, and $1.0\, \msun$
  (top panel) and $4.6\, \rm{Gyr}$ isochrones (bottom panel) for the
  following models: BCAH98 $\alpha=1.9$ (red, continuous), BCAH98
  $\alpha=1.0$ (red, dashed), SDF00 (blue), DAM97 (green) and DCJ08
  (cyan). Note that for masses $<~0.6\, \msun$ both computations of the BCAH98
  models are identical, this is due to the insensitivity to the
  value of the mixing length parameter in this mass regime.}
\label{fig:pre-ms_interior_models}
\end{figure}

Fig.~\ref{fig:pre-ms_interior_models} shows the variation between the
different pre-MS evolutionary tracks and the $4.6\, \rm{Gyr}$
isochrone for a range of masses. There are systematic
differences in both the mass tracks, especially at young ages, and
location of the MS, particularly for low-mass stars. These
differences stem from variations in the
treatment of various physical processes in addition to the values of
adopted parameters. Most notable are the treatment of convection, the
opacity sources and the treatment of the stellar interior/atmosphere
boundary (see \citealp{Hillenbrand04}).

The BCAH98 models have been
computed for two different mixing length parameters $\alpha=1.0$
(general mixing length parameter) and
$\alpha=1.9$ (solar calibrated value), where for stellar masses below $0.6\, \msun$ these models are
identical as the dependence on the mixing length parameter is
negligible in this regime \citep{Baraffe02}. Further note that the
publicly available SDF00 models have a slightly modified solar
abundance and mixing length parameter from the solar calibrated model
(cf. $Z=0.02$ and $\alpha=1.5$ with $Z=0.0249$ and $\alpha=1.605$ in
\citealp{Siess00}).

\subsection{Atmospheric models}
\label{atmospheric_models}

We required atmospheric model flux distributions to calculate a
theoretical colour-$T_{\rm{eff}}$ relation and bolometric corrections.
We produced a spectral library consisting of the \textsc{phoenix} BT-Settl model atmospheres for
$400 \leq T_{\rm{eff}} \leq 7800\, \rm{K}$ and the
\textsc{atlas9} models with newly updated opacity
distribution functions (ODFnew) for $8000 \leq T_{\rm{eff}} \leq
50\, 000\, \rm{K}$. Despite the differences in
microphysics between the two sets of atmospheric models, we found that
at the transitional $T_{\rm{eff}}=8000\, \rm{K}$ derived colours
from both sets of models agree to within $0.02\, \rm{mag}$ in all colours.

\subsubsection{The \textsc{atlas9}/ODFnew models}
\label{atlas9}

The \textsc{atlas9}/ODFnew
atmospheric models are those of
\cite{Castelli04}\footnote[2]{\url{ftp://ftp.stsci.edu/cdbs/grid/ck04models}}. These
models are based on the assumption
of steady-state plane-parallel layers under the influence of local
thermodynamic equilibrium and computed using the solar abundances of
\cite{Grevesse98}. Line blanketing effects are computed
statistically via the use of the opacity distribution functions
which average the contribution from the various atomic and molecular species
as described in \cite{Kurucz79}. A pure mixing length theory
\citep{Boehm58} is used in the computation of these models and we
adopted those termed ``no-overshoot'' on the basis that these better recreate the
observed spectra of stars with $T_{\rm{eff}}$ greater than the Sun
\citep*{Castelli97}. The mixing length parameter is set
to $\alpha = 1.25$ with a microturbulent velocity $\xi = 2\,
\rm{km\, s^{-1}}$.

\subsubsection{The BT-Settl models}
\label{phoenix}

The BT-Settl atmospheric models are those described in
\cite*{Allard11}\footnote[3]{\url{http://www.phoenix.ens-lyon.fr/Grids/BT-Settl/}}. These
models have been computed using an updated version of the
\textsc{phoenix} atmospheric code (cf. \citealp{Allard01}) to include
the updated \cite{Barber06} BT2 H$_{2}$O line list, the revised solar
abundances of \cite{Asplund09} and a sophisticated cloud model that
accounts for the settling of dust grains \citep{Allard03}. The effects
of turbulent mixing in the atmosphere are calculated by interpolating from 2-D and
3-D radiation hydrodynamic models \citep*{Ludwig06}, with the treatment
of dust for cooler models $(T_{\rm{eff}} < 2600\, \rm{K})$
based on the dust formation models of \cite{Freytag10}. The effects of
opacity are sampled directly using a library of over 700 million lines
including atomic and molecular species along the spectrum. The
BT-Settl models are computed under
the assumption of plane-parallel radiative transfer, where convection
is treated using the standard mixing length theory. The mixing length
parameter is defined to be $\alpha = 2.0$ with a microturbulent
velocity of $\xi = 2\, \rm{km\, s^{-1}}$.

\section{The Data}
\label{the_data}

\subsection{Observations}
\label{observations}

Our observations are a subset of a large survey of pre-MS clusters
(the remainder of which will be published in an upcoming paper; Bell et al. in prep.) and
were obtained using the 2.5-m \textit{Isaac Newton}
Telescope (INT) on La Palma. The survey was split over two
runs, using the same instrumentation and filter set on both occasions, namely the four EEV 2k
$\times$ 4k CCD Wide-Field Camera (WFC) with a
$34' \times\, 34'$ field-of-view and the $(UgriZ) _{_{\rm{WFC}}}$
filter set. A combination of long and
short exposures were used to ensure that any bright stars
saturated in the longer exposures could be measured in the
shorter exposures.  Whilst the $(gri)_{_{\rm{WFC}}}$ filters are based on
the Sloan Digital Sky Survey (SDSS) design there are no similar $u$
and $z$ filters for the WFC and so 
the Royal Greenwich Observatory (RGO) $U_{_{\rm{WFC}}}$ and $Z_{_{\rm{WFC}}}$ filters were
used. The first set of
observations were taken during the 21$^{\rm{st}}$ -- 26$^{\rm{th}}$ Oct. 2007 with the
second run during the 13$^{\rm{th}}$ -- 19$^{\rm{th}}$ Sept. 2008.

\begin{figure}
\centering
\includegraphics[width=\columnwidth]{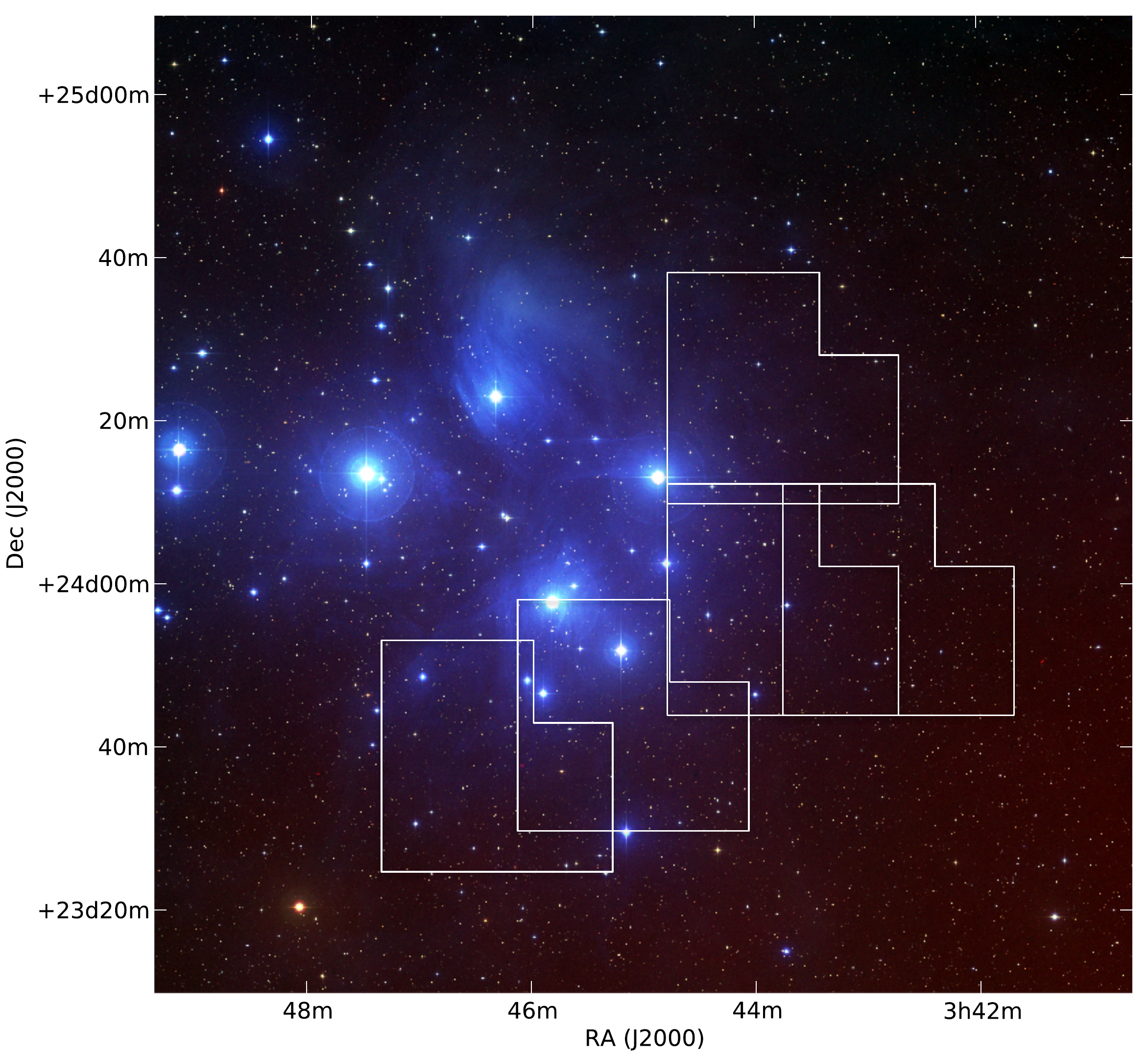}
\caption{Two square degree three-colour mosaic image of the Pleiades
  from the ESO Digital Sky Survey with the five INT-WFC fields-of-view overlaid. Note that the image is
  slightly displaced from the central coordinates of $\alpha = 03^{\rm{h}}\, 47^{\rm{m}}\,
  24^{\rm{s}}, \delta = +24^\circ\, 07'\, 00''$ to illustrate the region chosen to
  identify low-mass members.}
\label{fig:pleiades_mosaic}
\end{figure}

\begin{table}
\caption{The central coordinates for each field-of-view and
  exposure times in the INT-WFC bandpasses for the observations of the
  Pleiades.}
\vspace{0.05in}
\begin{tabular}{c c c c c}
\hline  \hline
Field&RA Dec.   &Filter&Exposure time (s)\\
       &(J2000.0)&         &$\times\, 1$ unless stated \\
\hline
Pleiades&$03^{\rm{h}}\,46^{\rm{m}}\,19.2^{\rm{s}}$&$U_{_{\rm{WFC}}}$&1, 10\\
Field~A&$+23^\circ\,38'\,02.4''$&$g_{_{\rm{WFC}}}$&1, 10, 100, 500($\times\,2$)\\
&&$r_{_{\rm{WFC}}}$&1, 7, 50, 250\\
&&$i_{_{\rm{WFC}}}$&1, 10, 100($\times\,2$)\\
&&$Z_{_{\rm{WFC}}}$&1, 7, 50, 250\\
\hline
Pleiades&$03^{\rm{h}}\,45^{\rm{m}}\,00.0^{\rm{s}}$&$U_{_{\rm{WFC}}}$&1, 10\\
Field~B&$+23^\circ\,44'\,45.6''$&$g_{_{\rm{WFC}}}$&1, 10, 100, 500($\times\,2$)\\
&&$r_{_{\rm{WFC}}}$&1, 7, 50, 250\\
&&$i_{_{\rm{WFC}}}$&1, 10, 100($\times\,2$)\\
&&$Z_{_{\rm{WFC}}}$&1, 7, 50, 250\\
\hline
Pleiades&$03^{\rm{h}}\,43^{\rm{m}}\,44.4^{\rm{s}}$&$U_{_{\rm{WFC}}}$&1, 10\\
Field~C&$+23^\circ\,58'\,08.4''$&$g_{_{\rm{WFC}}}$&1, 10, 100, 500($\times\,2$)\\
&&$r_{_{\rm{WFC}}}$&1, 7, 50, 250\\
&&$i_{_{\rm{WFC}}}$&1, 10, 100($\times\,2$)\\
&&$Z_{_{\rm{WFC}}}$&1, 7, 50, 250\\
\hline
Pleiades&$03^{\rm{h}}\,42^{\rm{m}}\,36.0^{\rm{s}}$&$U_{_{\rm{WFC}}}$&1, 10\\
Field~D&$+23^\circ\,58'\,08.4''$&$g_{_{\rm{WFC}}}$&1, 10, 100, 500($\times\,2$)\\
&&$r_{_{\rm{WFC}}}$&1, 7, 50, 250\\
&&$i_{_{\rm{WFC}}}$&1, 10, 100($\times\,2$)\\
&&$Z_{_{\rm{WFC}}}$&1, 7, 50, 250\\
\hline
Pleiades&$03^{\rm{h}}\,43^{\rm{m}}\,44.4^{\rm{s}}$&$U_{_{\rm{WFC}}}$&1, 10\\
Field~E&$+24^\circ\,24'\,28.8''$&$g_{_{\rm{WFC}}}$&1, 10, 100, 500($\times\,2$)\\
&&$r_{_{\rm{WFC}}}$&1, 7, 50, 250\\
&&$i_{_{\rm{WFC}}}$&1, 10, 100($\times\,2$)\\
&&$Z_{_{\rm{WFC}}}$&1, 7, 50, 250\\
\hline \hline
\end{tabular}
\label{tab:pleiades_exp_times}
\end{table}

Five overlapping fields within the Pleiades open cluster were observed over the
course of the two runs (see Fig.~\ref{fig:pleiades_mosaic} and
Table~\ref{tab:pleiades_exp_times}). In addition, four fields in
Stripe~82 of the
SDSS were used as standard fields and routinely observed throughout
both runs. 

\subsection{Data reduction}
\label{data_reduction}

Each image was debiased and flat fielded using a median stacked bias and
flat field frames. Both
the $i_{_{\rm{WFC}}}$- and $Z_{_{\rm{WFC}}}$-band images were defringed using library
fringe frames. The data were analysed using the \textsc{cluster}
package as described in \cite{Naylor02}, \cite{Burningham03} and
\cite{Jeffries04}.
The $i_{_{\rm{WFC}}}$-band images for a particular field were combined
(after calculating spatial transformations between separate frames)
with the resulting deep image used for object identification and
detection. An object list was created and optimal photometry \citep{Naylor98}
was then performed providing a flux for each
object at a given position. A spatially varying profile correction was
calculated from the bright unsaturated stars in each frame so that the
photometry could later be calibrated using standard star
observations. Although the $i_{_{\rm{WFC}}}$-band images
were combined, optimal photometry was carried out on each individual
frame, so as to ensure that the good signal-to-noise in one frame was
not outweighed by a poorer signal-to-noise in another frame upon
combining the images. The individual profile corrected measurements of each frame were
then combined by weighting each measurement in accordance with its
signal-to-noise ratio and accounting for differences in airmass
between separate frames. An additional statistical uncertainty was
added to each measurement at this stage to ensure that the
distribution of chi-squared $(\chi^{2})$ resulting from combining the measurements
 versus signal-to-noise was independent of
signal-to-noise and of
order unity \citep{Naylor02}. The additional statistical uncertainty
adopted here ranged from $0.01-0.03\, \rm{mag}$ and reflects uncertainties in the profile
correction. Stars with a reduced $\chi^{2} > 10$ are generally flagged
as variable (see \citealp{Burningham03}). For some fields-of-view, we have combined photometric
observations at two different epochs and due to pre-MS
variability over such timescales, there are a number of
sources that have thus been flagged. For sources which displayed
night-to-night variations, we have replaced the photometry with
colours and magnitudes from a single night. The Two-Micron All-Sky Survey
(2MASS, \citealp{Cutri03}) was used to provide an astrometric solution for each frame.
The rms residual of the
6-coefficient fit were approximately $0.2\, \rm{arcsec}$.

\begin{table*}
\caption{A sample of the full Pleiades photometric catalogue with colours and magnitudes in the
  natural INT-WFC photometric system. Due to space restrictions, we only show
  the $g_{_{\rm{WFC}}}$ and $(g-i)_{_{\rm{WFC}}}$ colours and
  magnitudes as a representation of its content. The full photometric
  catalogue (available online) also includes photometry in the
  $(U-g)_{_{\rm{WFC}}}$, $(r-i)_{_{\rm{WFC}}}$, and
  $(i-Z)_{_{\rm{WFC}}}$ colours and the $U_{_{\rm{WFC}}}$,
  $r_{_{\rm{WFC}}}$, $i_{_{\rm{WFC}}}$, and $Z_{_{\rm{WFC}}}$ magnitudes.
  Columns list unique
  identifiers for each star in the catalogue: field and CCD number
  (integer and decimal), ID, RA and
  Dec. (J2000.0), CCD pixel coordinates of the star, and for each of
  $g_{_{\rm{WFC}}}$ and $(g-i)_{_{\rm{WFC}}}$ there is a magnitude, an
  uncertainty on the magnitude and a flag (OO represents a ``clean
  detection''; see \protect\citealp{Burningham03} for a full description of
  the flags.)}
\begin{tabular}{c c c c c c c c c c c c}
\hline \hline
Field&ID&RA (J2000.0)&Dec. (J2000.0)&x&y&$g_{_{\rm{WFC}}}$&$\sigma_{
g_{_{\rm{WFC}}}}$&Flag&$(g-i)_{_{\rm{WFC}}}$&$\sigma_{
(g-i)_{_{\rm{WFC}}}}$&Flag\\
\hline
21.04&8&03 46 34.136&+23 37
26.25&1204.056&1439.980&8.236&0.010&OO&-0.412&0.014&OL\\
22.02&16&03 43 41.521&+23 38
56.30&998.324&2035.800&8.523&0.010&OS&-2.653&0.014&SS\\
\hline \hline
\end{tabular}
\label{tab:pleiades_photometry}
\end{table*}

To create a final optical catalogue of the Pleiades based on the five
overlapping fields, it was necessary to combine the individual
fields-of-view. For this process, we used the normalisation procedure
as described in \cite{Jeffries04}. A running
catalogue was created that contained
all the derived magnitudes and colours for each object identified in
all fields. We then calculated the mean magnitude and colour
differences of stars in common between two overlapping fields,
and adjusted the zero-points for each field to minimise these
differences (in effect allowing for small variations in the zero-point
between fields observed on different nights). This process ensures greater consistency between fields
in the final catalogue. The resulting zero-point shift is an indicator of the
internal consistency of the photometry, as well as the accuracy with
which the profile corrections was performed, and suggests an accuracy
of $\simeq 1$~per~cent. Our full Pleiades photometric catalogue is
given in Table~\ref{tab:pleiades_photometry}.

\subsection{Zero-point stability}
\label{zero-point_stability}

Standard field observations were taken on five separate
evenings. On only one of the five nights were conditions photometric
throughout, the other four nights were affected either at the beginning or
the end by the presence of cloud. 
We therefore calculated a zero point for each $g_{_{\rm{WFC}}}$-band observation by 
comparing the instrumental magnitudes to those in the SDSS catalogue,
and correcting for extinction using a mean $g_{_{\rm{WFC}}}$-band extinction coefficient 
($k_{g_{_{\rm{WFC}}}} =0.19$) for La Palma.
We could then identify when the zero-point became unstable and define
half nights of bona fide photometric conditions, which were then used for the
photometric calibration procedure (the maximum deviation over the course of a
single half night was $0.05\, \rm{mag}$). Due to a paucity of blue stars in our
standards catalogue, we opted to use the colour range which was most
populated ($0.6 \leq (g-i)_{_{\rm{WFC}}} \leq 1.0$) and define our zero-point at a
median colour of $(g-i)_{_{\rm{WFC}}} = 0.8$. This colour range cut in $(g-i)_{_{\rm{WFC}}}$
defines the necessary cuts in the other INT-WFC colours ($1.05 \leq (U-g)
_{_{\rm{WFC}}} \leq 1.96$, $0.14 \leq (r-i)_{_{\rm{WFC}}}
\leq 0.31$ and $0.02 \leq (i-Z) _{_{\rm{WFC}}} \leq 0.18$ with median values of
$(U-g) _{_{\rm{WFC}}} = 1.505$, $(r-i)_{_{\rm{WFC}}} = 0.225$ and
$(i-Z) _{_{\rm{WFC}}} = 0.1$). The seeing
generally varied between $1$ and $2\, \rm{arcsec}$ in the temporal regions
defined above.

\subsection{Transforming the data into the standard system}
\label{photometric_calibration}

\begin{figure}
\centering
\includegraphics[width=\columnwidth]{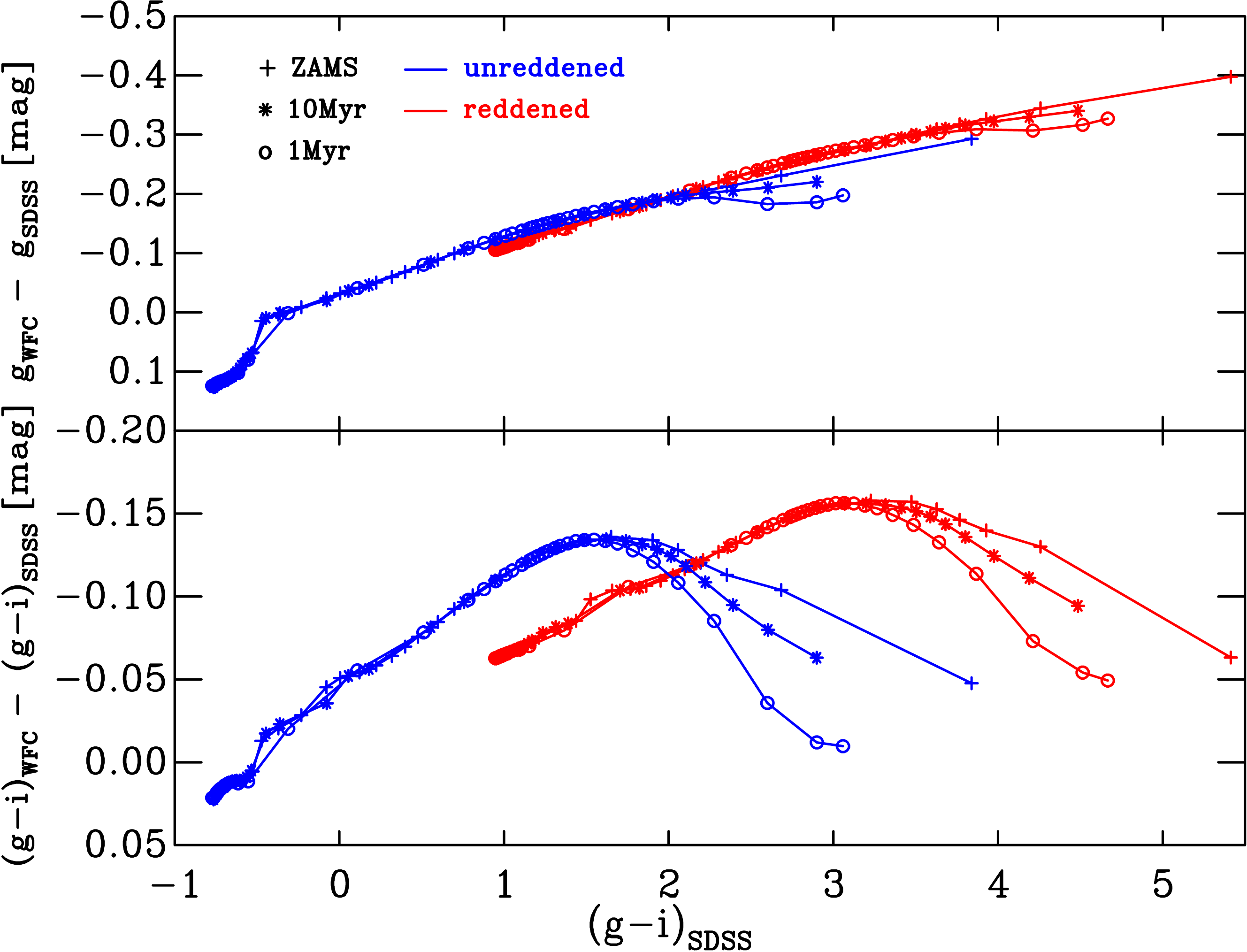}
\caption{Calculated transformations between the SDSS and INT-WFC
  photometric systems. The transformations have been calculated for
  reddened (nominal $E(B-V) = 1\, \rm{mag}$; red) and
  unreddened (blue) ZAMS (crosses), $10\, \rm{Myr}$ (asterisks) and $1\,
  \rm{Myr}$ pre-MS stars (open circles). Upper panel: $\Delta g$ versus
  $(g-i)_{_{\rm{SDSS}}}$. Lower
  panel: $\Delta (g-i)$ versus $(g-i)_{_{\rm{SDSS}}}$. No single
  (even non-linear) transformation will calibrate observations of both
  red ($(g-i)_{_{\rm{SDSS}}} \gtrsim 1.8$) MS and pre-MS stars into a
  standard system, and if either are reddened the situation becomes worse
  ($\simeq 0.15\, \rm{mag}$ at a colour $(g-i)_{_{\rm{SDSS}}} \simeq 3$).}
\label{fig:delta_wfc_sdss_red}
\end{figure}

Traditional photometric calibration transforms instrumental magnitudes 
into magnitudes on a standard system.
Typically colours and magnitudes are transformed using linear
functions of the form

\begin{equation}
g_{\rm{stand}} = \psi_{g}(g-i)_{\rm{inst}} - k_{g}\chi + z_{g},
\label{trad_zp_g}
\end{equation}

\begin{equation}
(g-i)_{\rm{stand}} = \psi_{gi}(g-i)_{\rm{inst}} - k_{gi}\chi + z_{gi},
\label{trad_zp_gi}
\end{equation}

\noindent where $\chi$ is the airmass, $k$ the extinction
coefficients, $z$ the zero-points and $\psi$ the colour
terms ($k$, $z$ and $\psi$ are determined from observations of
standard stars).
Fig.~\ref{fig:delta_wfc_sdss_red} shows the $g$-band magnitude and
$g-i$ colour
differences between the standard SDSS and the natural INT-WFC
photometric systems (see Appendices 
\ref{creating_int/wfc_system_responses} and
\ref{relative_differences_synthetic_photometry}
for details).
It is immediately obvious that there is
no linear colour-dependent transformation that can be applied
to INT-WFC observations of MS stars to calibrate them into
the standard SDSS photometric system.
Even if non-linear MS transformations were created, it is clear that they could not
be used to transform observations of red ($(g-i)_{\rm{SDSS}} \gtrsim 1.8$) pre-MS stars into the standard system.
This is due to differences in the spectra between MS and pre-MS stars
of the same colour.
These differences can result in both $g$-band
magnitude and $g-i$ colour
errors of order $0.1-0.15\, \rm{mag}$ at a colour of $(g-i)_{_{\rm{SDSS}}}
\simeq 3$, culminating in stars that ultimately occupy the wrong
position in CMD space. Even larger errors can be caused by the difference between the spectrum of a
reddened and an unreddened star of the same apparent colour (see
Fig.~\ref{fig:delta_wfc_sdss_red}).
From the calculated SDF00 zero-age main-sequence
(ZAMS; see Appendix~\ref{relative_differences_synthetic_photometry})
the differential behaviour between MS and pre-MS transformations
occurs at an approximate spectral type of between K6 and M0.
\cite{Mayne12} document a similar, though more extreme situation which can
result from using MS transformations for brown-dwarfs.

A deviation of up to $0.15\, \rm{mag}$ may
help explain, to some extent, why pre-MS
isochrones do not simultaneously fit both the higher and lower mass
members within a given cluster
\citep{Hartmann03,Stauffer07,Jeffries09}, although other factors
such as photometric variability, extinction, unresolved binaries,
accretion luminosity and poorly constrained opacity line lists in the
atmospheric models may also contribute to this
\citep{Hartmann01,Burningham05a}. Furthermore, this could also have
severe implications for cluster age and mass function
determinations. Allowing the pre-MS stars in a CMD to ``float'' by $0.15\,
\rm{mag}$ in both colour and magnitude can result in an age difference of a
factor of two through isochrone fitting. 
The conversion from observable magnitudes and colours into
mass estimates are highly age and model dependent. Hence an
incorrrect age would result in an erroneous mass function estimate
thus making meaningful comparison between cluster mass functions
difficult at best.

\subsection{Calibrating the natural system}
\label{calibration_int/wfc_photometry}

\begin{figure}
\centering
\includegraphics[width=\columnwidth]{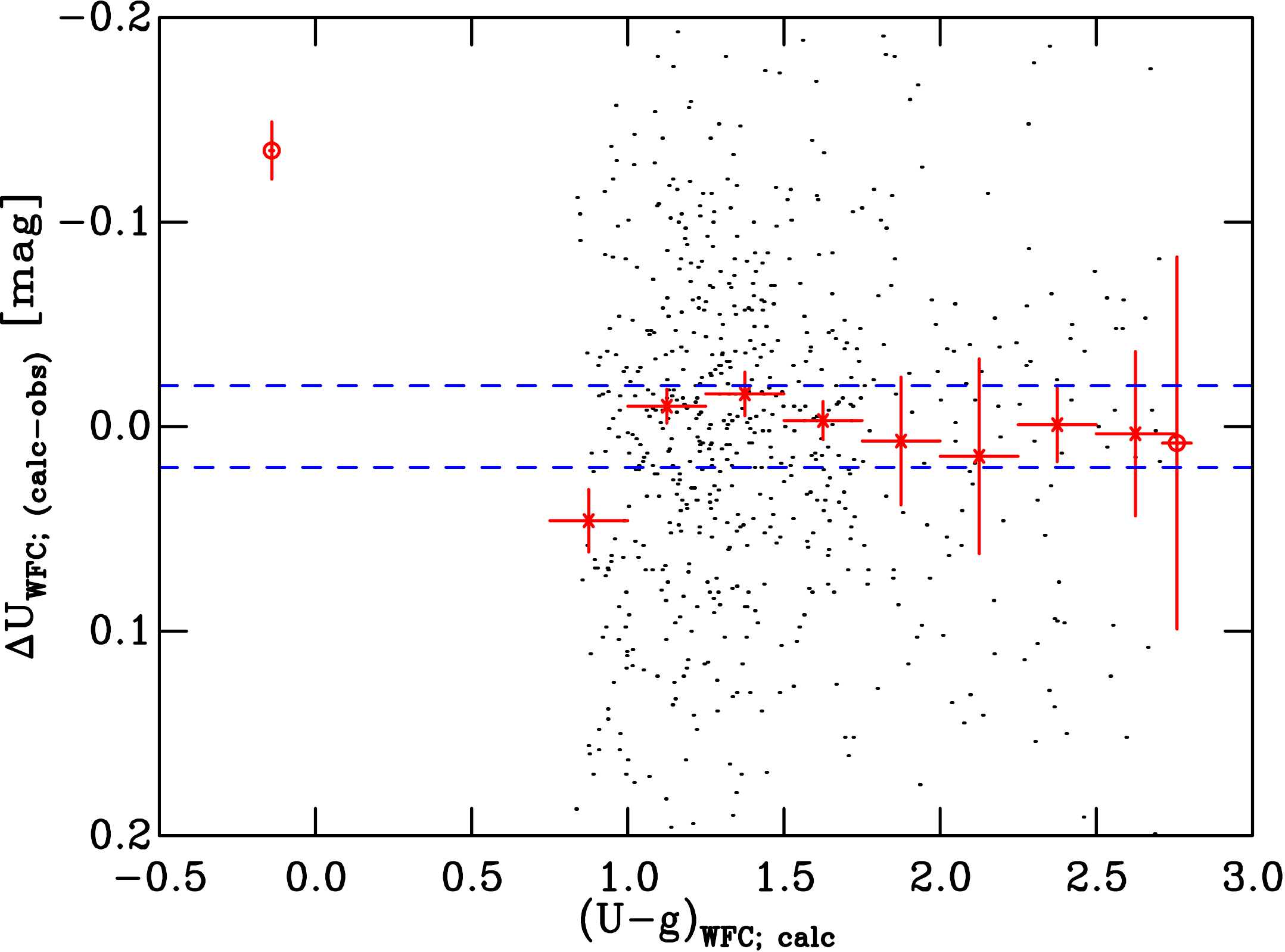}
\caption{Magnitude difference in the $U_{_{\rm{WFC}}}$-band between the
  $\rm{WFC_{calc}}$ and $\rm{WFC_{obs}}$ photometric
  catalogues as a function of $(U-g)_{_{\rm{WFC; calc}}}$. Asterisks represent the median value of all points in a
  given bin of size $0.25\, \rm{mag}$ in colour. The error bars on these
  symbols are the standard error about the median value. Open circles represent
  individual data points and are plotted where the number of points is
  insufficient to calculate the median value in a given bin (defined
  as five). The error
  bars of these symbols are the individual uncertainties associated
  with that point. The dashed lines represents the $\pm\, 0.02\, \rm{mag}$
  level with respect to zero.}
\label{fig:final_delta_wfc_u}
\end{figure}

\begin{figure}
\centering
\includegraphics[width=\columnwidth]{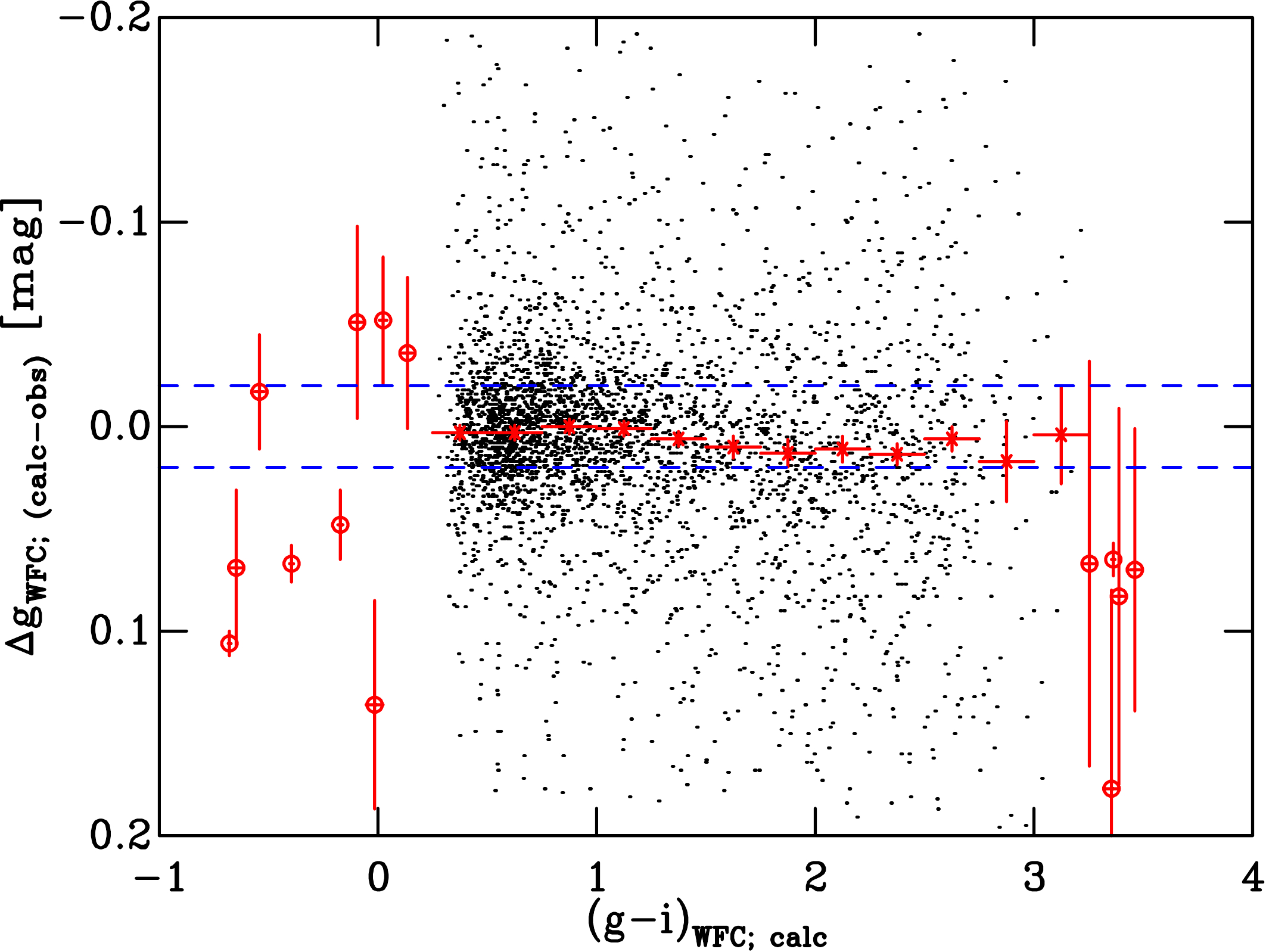}
\caption{Same as Fig.~\ref{fig:final_delta_wfc_u} but for the
  difference in the $g_{_{\rm{WFC}}}$-band as a function of
  $(g-i)_{_{\rm{WFC; calc}}}$.}
\label{fig:final_delta_wfc_g}
\end{figure}

\begin{figure}
\centering
\includegraphics[width=\columnwidth]{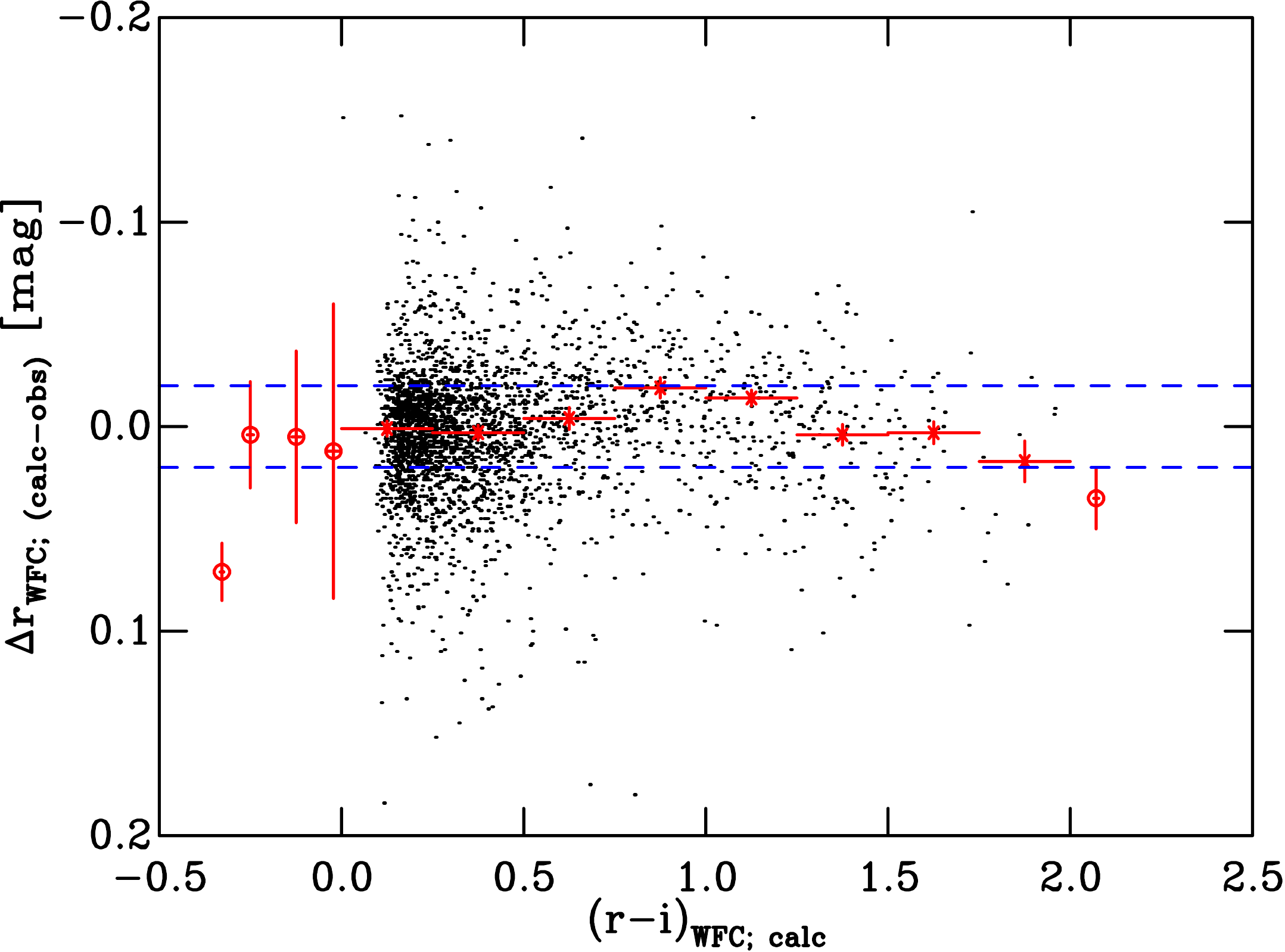}
\caption{Same as Fig.~\ref{fig:final_delta_wfc_u} but for the
  difference in the $r_{_{\rm{WFC}}}$-band as a function of
  $(r-i)_{_{\rm{WFC; calc}}}$.}
\label{fig:final_delta_wfc_r}
\end{figure}

\begin{figure}
\centering
\includegraphics[width=\columnwidth]{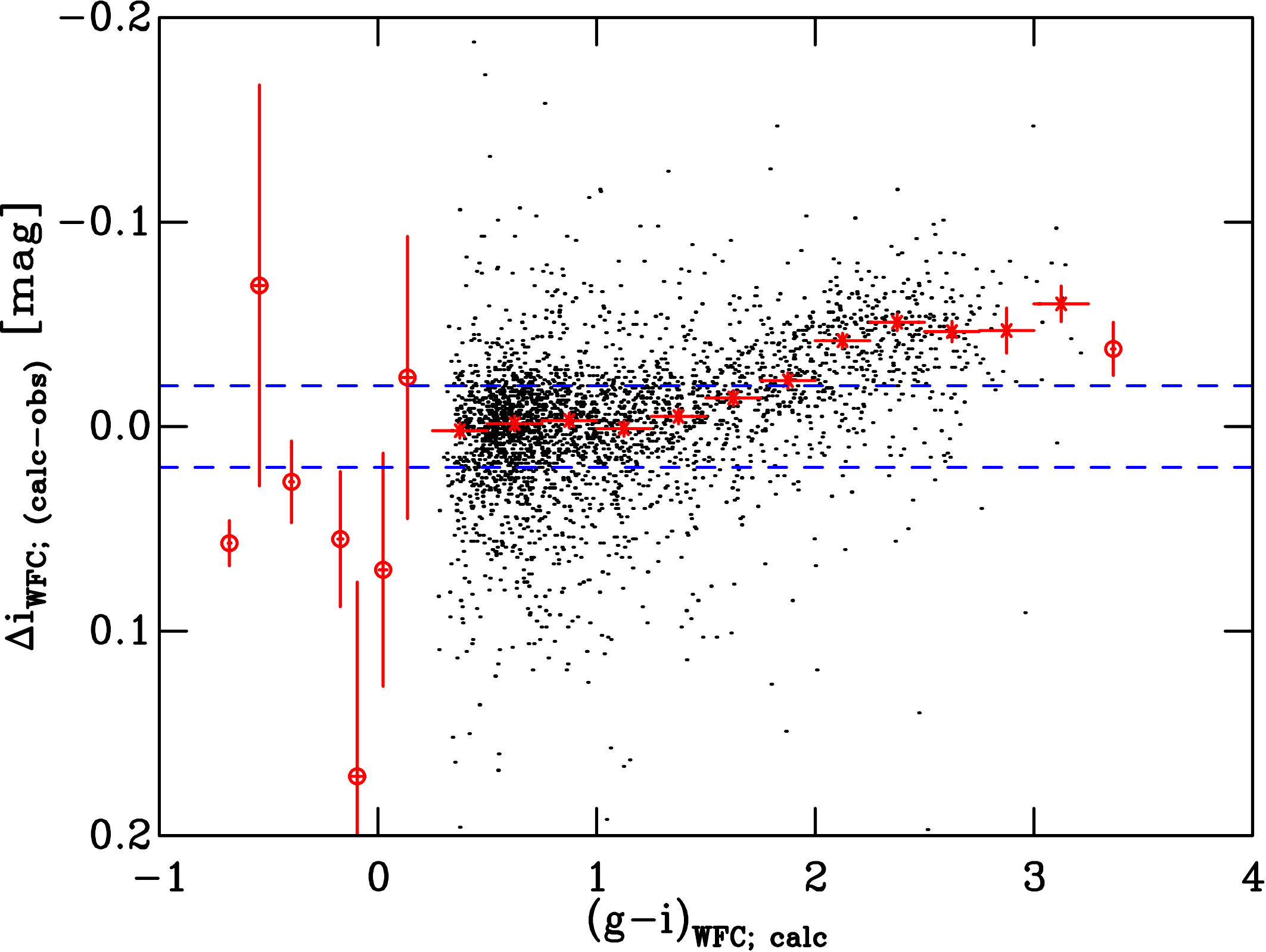}
\caption{Same as Fig.~\ref{fig:final_delta_wfc_u} but for the
  difference in the $i_{_{\rm{WFC}}}$-band as a function of
  $(g-i)_{_{\rm{WFC; calc}}}$.}
\label{fig:final_delta_wfc_i}
\end{figure}

\begin{figure}
\centering
\includegraphics[width=\columnwidth]{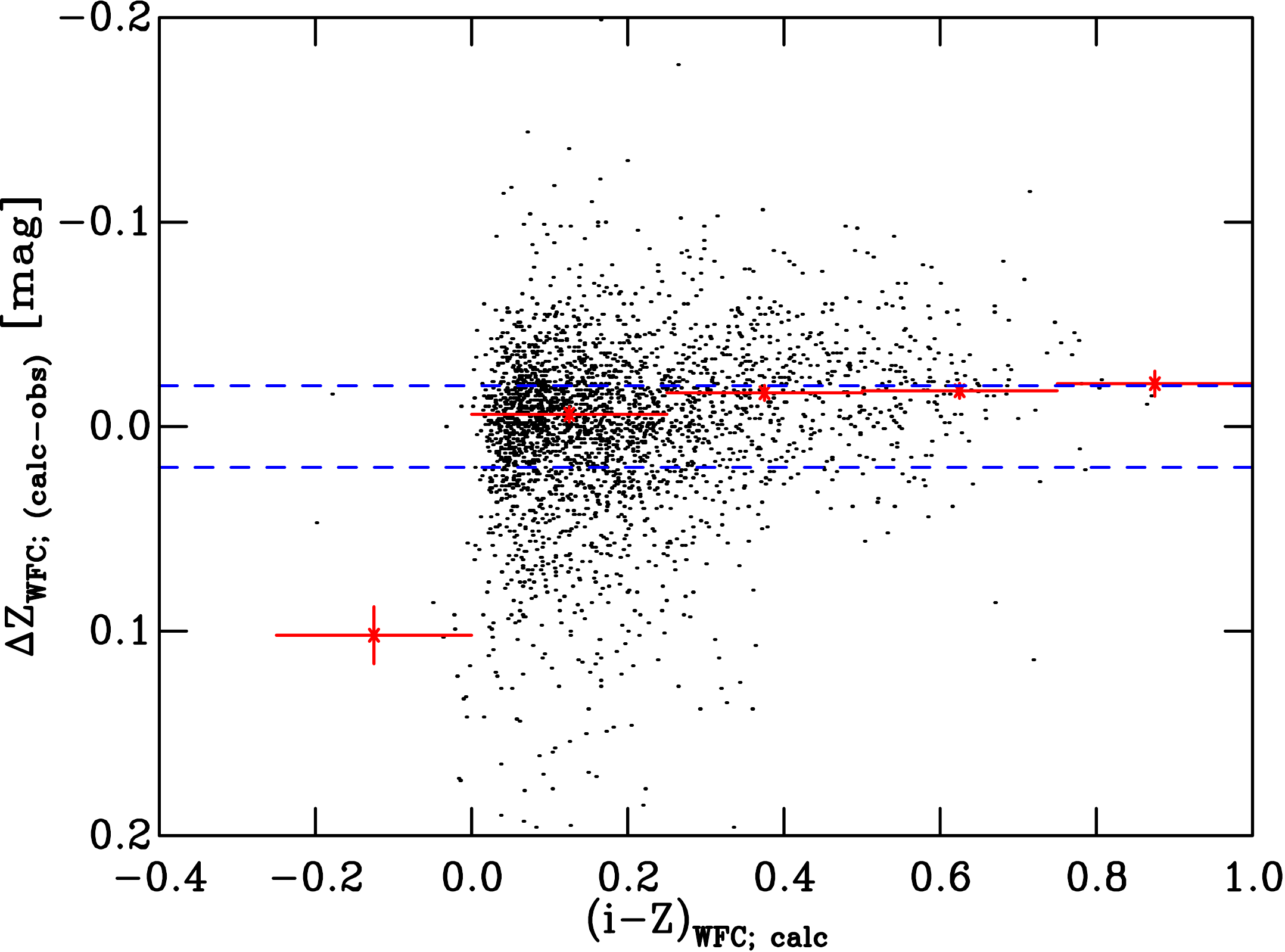}
\caption{Same as Fig.~\ref{fig:final_delta_wfc_u} but for the
  difference in the $Z_{_{\rm{WFC}}}$-band as a function of
  $(i-Z)_{_{\rm{WFC; calc}}}$.}
\label{fig:final_delta_wfc_z}
\end{figure}

If the observations cannot be transformed into a
standard photometric system, then the photometry must remain in
the natural photometric system and the
theoretical models transformed into this system.
To achieve this we need a reliable model for the throughput 
for each bandpass as a function of wavelength.
We emphasise that this is not simply a matter of using the filter responses, or even
multiplying the CCD quantum efficiency by the filter response, but involves allowing 
for every element from the Earth's atmosphere through to the CCD to create a \textit{system}
response. We present our best estimates for the system responses in
Appendix~\ref{creating_int/wfc_system_responses}.

Having obtained the system responses it is crucial that we obtain an estimate of how
precise these are. In principle we could achieve this by
folding the observed fluxes of spectrophotometric standards through our estimated
system responses and seeing how closely the observed and predicted magnitudes match.
In practice such standards have not been measured in sufficient numbers with the range 
of colours and magnitudes to prove useful for our purposes.
Instead we took a sample of MS stars, used their observed SDSS colours to establish 
their effective temperatures, and then folded model atmospheres of the appropriate 
temperature through our estimated system responses to find their magnitudes in 
the INT-WFC system.

\begin{table*}
\caption{The calculated transformations between the SDSS and INT-WFC
  photometric systems for unreddened MS stars where
  $\Delta$ represents (INT-WFC $-$ SDSS). The full table
  is available online, a sample is shown here as a
  representation of its content.}
\begin{tabular}{c c c c c c c c c c c}
\hline \hline
$T_{\rm{eff}}$&log$\,g$&$\Delta
g$&$\Delta(g-i)$&$(g-i)_{_{\rm{SDSS}}}$&$\Delta(u-g)$&$(u-g)_{_{\rm{SDSS}}}$&$\Delta(r-i)$&$(r-i)_{_{\rm{SDSS}}}$&$\Delta(i-z)$&$(i-z)_{_{\rm{SDSS}}}$\\
\hline
2775&5.25&-0.29300&-0.04765&3.83961&0.01264&4.12961&0.24688&2.20001&-0.17476&1.22123\\
3290&5.11&-0.23094&-0.10391&2.68386&-0.02563&3.06419&0.12898&1.20809&-0.08704&0.66867\\
3528&4.98&-0.21237&-0.11299&2.35355&-0.03358&2.87477&0.09764&0.94887&-0.06707&0.53108\\
3683&4.92&-0.20428&-0.12031&2.19344&-0.03595&2.83717&0.07940&0.81597&-0.05550&0.45615\\
3829&4.87&-0.19761&-0.12806&2.05653&-0.03751&2.81937&0.06188&0.70141&-0.04429&0.38859\\
\hline \hline
\end{tabular}
\label{tab:main-sequence_transformations}
\end{table*}

We defined our sample of MS stars using a series of colour-colour
diagrams to isolate the observed MS in the Stripe~82 fields and trim
any obvious outliers.
To obtain colours and magnitudes in the INT-WFC system we then transformed the 
standard SDSS colours and magnitudes into the INT-WFC photometric system (hereafter
$\rm{WFC_{calc}}$).
These transforms are calculated by folding model atmospheres through both the
SDSS and our estimated INT-WFC responses and are given in
Table~\ref{tab:main-sequence_transformations}.
The details of these calculations are given in
Appendix~\ref{relative_differences_synthetic_photometry}, and the
transformed standards in the INT-WFC colours are given in
Table~\ref{tab:standards_int-wfc_photometry}. 

\begin{table*}
\caption{A sample of the transformed Stripe~82 standard star catalogue with colours and magnitudes in the
  natural INT-WFC photometric system. The full
  photometric (available online) also includes photometry in the
  $(U-g)_{_{\rm{WFC}}}$, $(r-i)_{_{\rm{WFC}}}$, and
  $(i-Z)_{_{\rm{WFC}}}$ colours. Columns list unique identifiers for
  each star in the catalogue (see Table~\ref{tab:pleiades_photometry})
  for details.}
\begin{tabular}{c c c c c c c c c c c c}
\hline \hline
Field&ID&RA (J2000.0)&Dec. (J2000.0)&x&y&$g_{_{\rm{WFC}}}$&$\sigma_{
g_{_{\rm{WFC}}}}$&Flag&$(g-i)_{_{\rm{WFC}}}$&$\sigma_{
(g-i)_{_{\rm{WFC}}}}$&Flag\\
\hline
0.00&67737&20 47 07.346&-01 05
52.53&0.000&0.000&17.158&0.006&OO&0.585&0.008&OO\\
0.00&67740&20 47 09.412&-01 04
17.09&0.000&0.000&14.721&0.006&OO&0.482&0.008&OO\\
\hline \hline
\end{tabular}
\label{tab:standards_int-wfc_photometry}
\end{table*}

With the INT-WFC standards established, we can evaluate how well
constrained the calculated system responses of the INT-WFC are. Calculating
zero-points in the colour ranges specified in
Section~\ref{zero-point_stability} and applying these zero-points and
extinction coefficients only (no colour terms) to
all standard field observations we created a set of observed standard
star colours and magnitudes in the INT-WFC system. To create a single catalogue for each
of the four standard fields we performed the normalisation process as
described in Section~\ref{data_reduction} and merged all four
fields to create a catalogue of observed colours and magnitudes of Stripe~82 standards in the
INT-WFC photometric system (hereafter
$\rm{WFC_{obs}}$). After the normalisation procedure, we examined, for
each star,
the magnitude shift between each observation and the mean observed
magnitude. This is an indicator of the internal precision of our photometry
and we found that $g_{\rm{rms}} = 0.011\, \rm{mag}$, $(g-i)_{\rm{rms}} =
0.014\, \rm{mag}$, $(U-g)_{\rm{rms}} = 0.020\, \rm{mag}$, $(r-i)_{\rm{rms}}
= 0.011\, \rm{mag}$ and $(i-Z)_{\rm{rms}} = 0.010\, \rm{mag}$. We
attribute the slightly poorer precision in $U_{_{\rm{WFC}}}$ to
a lack of colour dependent atmospheric extinction terms.

The differences $\rm{WFC_{calc}} - \rm{WFC_{obs}}$ are shown in
Figs.~\ref{fig:final_delta_wfc_u}~--~\ref{fig:final_delta_wfc_z}. Using bin
sizes of $0.25\, \rm{mag}$ in colour, and only data with uncertainties of
less than $0.03\, \rm{mag}$ we calculated the median value in each
colour bin as an indicator of the level of agreement between the
$\rm{WFC_{calc}}$ and $\rm{WFC_{obs}}$ catalogues in a given
colour range. A paucity of data at extreme colours forced
us to relax our restrictions on the uncertainty in the data points to
less than $0.1\, \rm{mag}$ and where the number of stars was
insufficient to calculate the median value in a given colour bin (taken
as five), the individual data points were plotted.
These plots show that we reproduce the zero point to
within $0.01\, \rm{mag}$ at the colours which we defined to calculate
the zero-points in Section~\ref{zero-point_stability}.
The remaining residuals lie within the $\pm 0.02\, \rm{mag}$ level across the entire
colour range except for a small region in the $U_{_{\rm{WFC}}}$-band and colours
redder than $(g-i)_{_{\rm{WFC}}} \simeq 2$ in the $i_{_{\rm{WFC}}}$-band.

The most obvious causes of any disagreement between $\rm{WFC_{calc}}$ and 
$\rm{WFC_{obs}}$ are that our estimated system responses are incorrect, and
that the atmospheric models are incorrect.
In an attempt to obtain some idea of the likely magnitude of the latter effect we
tried re-calculating the transforms using different model atmospheres
(the \textsc{phoenix}/GAIA models of \citealp{Brott05}) and the
observed spectral library of \cite{Pickles98}.
We found these changed the transformations by up to $0.02\, \rm{mag}$, nicely explaining
the majority of the residuals.
It remains unclear whether the remaining residuals are due to 
errors in the atmospheric fluxes, or poorly modelled 
$U_{_{\rm{WFC}}}$- and $i_{_{\rm{WFC}}}$-band system responses. 
As a result, we retain the system responses in their current
calculated form for the remainder of this study, and note the fact there are 
uncertainties at the levels shown in Figs.~\ref{fig:final_delta_wfc_u}~--~\ref{fig:final_delta_wfc_z}. 

\section{Comparing the models and the data -- The Pleiades}
\label{testing_the_models_using_cmds}

\begin{table*}
\caption{A sample of the catalogue of colours and magnitudes in the
  natural INT-WFC photometric system for Pleiades members.
  The columns and content are in
  the same format as Table~\ref{tab:pleiades_photometry}. The full
  photometric catalogue is available online.}
\begin{tabular}{c c c c c c c c c c c c}
\hline \hline
Field&ID&RA (J2000.0)&Dec. (J2000.0)&x&y&$g_{_{\rm{WFC}}}$&$\sigma_{
g_{_{\rm{WFC}}}}$&Flag&$(g-i)_{_{\rm{WFC}}}$&$\sigma_{
(g-i)_{_{\rm{WFC}}}}$&Flag\\
\hline
24.02&45&03 41 03.006&+23 43
21.42&426.717&3696.221&17.653&0.006&OO&2.553&0.006&OO\\
24.02&118&03 41 05.230&+23 50
14.82&526.182&2444.212&20.400&0.015&OO&3.228&0.017&OO\\
\hline \hline
\end{tabular}
\label{tab:pleiades_members_photometry}
\end{table*}

Having characterised the natural INT-WFC photometric system and
calculated the corresponding system responses, we can now test the
pre-MS interior and stellar atmospheric models by
comparing them to our Pleiades data.
We obtained a list of approximately 180 members, by cross-correlating our
full catalogue with the combined membership catalogues of
\cite{Stauffer07} and \cite*{Lodieu12}. This membership list is given in
Table~\ref{tab:pleiades_members_photometry}.

\subsection{Model parameters}
\label{model_parameters}

We used a distance modulus $dm = 5.63\pm 0.02$ which is based on
the trigonometric parallax of \cite{Soderblom05}.
We chose to rely on the lithium depletion age, since although it is based 
on the predictions of stellar interior models, \cite{Jeffries05} show there 
is a very high level of agreement between the various models.
Furthermore we used the age based on the boundary in $K_{\rm{s}}$ ($130\, \rm{Myr}$) of 
\cite*{Barrado04b}, since as we shall show later,  $K_{\rm{s}}$ is the best 
predictor of luminosity available to us.
We used a reddening of $E(B-V)=0.04$ based on the mean extinction $A_{V}=0.12$ 
(see \citealp{Stauffer98a}) and $R_{V}=3.2$.
Hence, to transform the theoretical models into the observable
plane, we first reddened the atmospheric models
by a nominal $E(B-V)=0.04$ using the reddening law of
\cite*{Cardelli89} and then calculated bolometric corrections using
Eqn.~\ref{bc_final}.

\begin{figure*}
\centering
\includegraphics[width=0.9\textwidth]{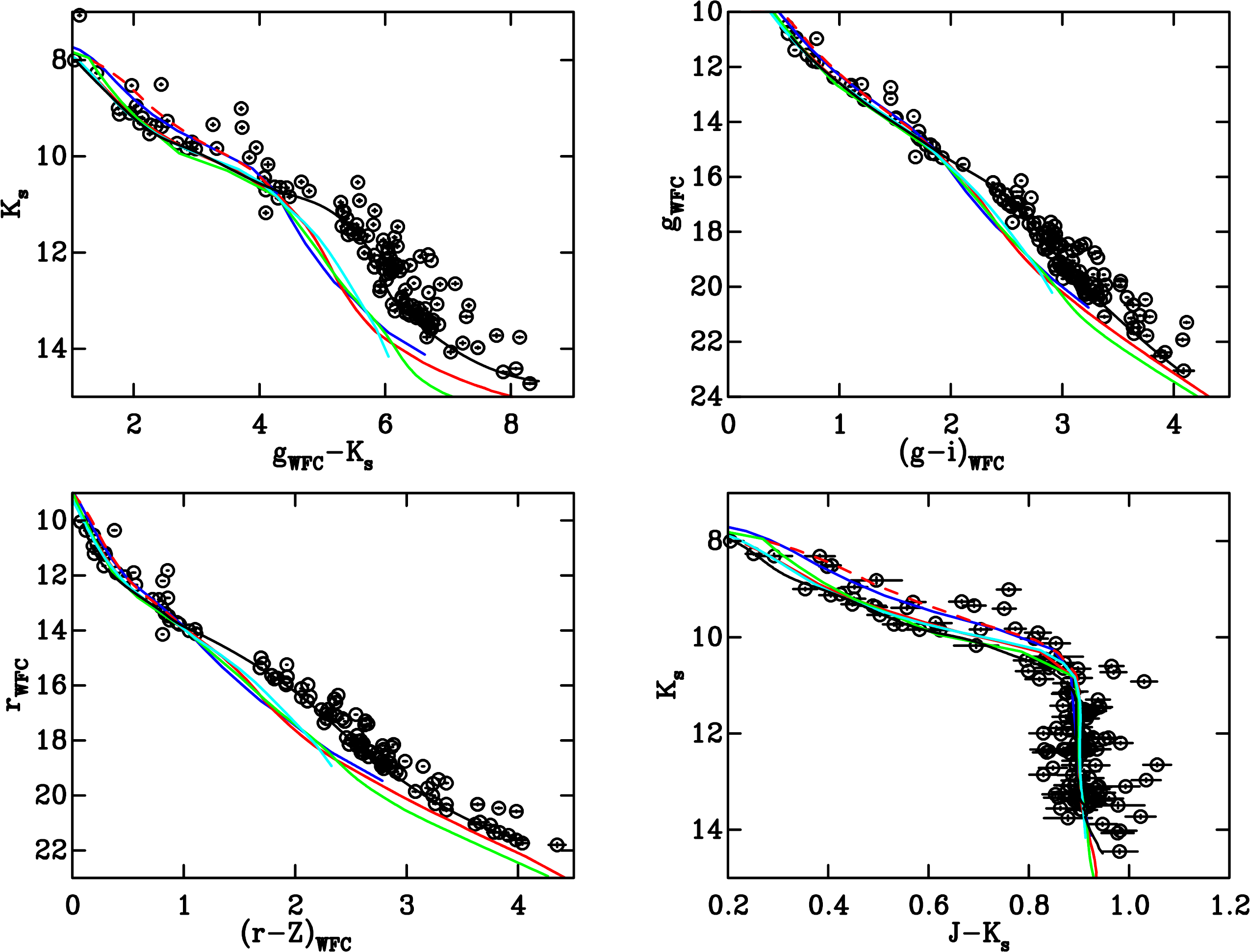}
\caption{Optical/near-infrared CMDs of Pleiades members. The $130\, \rm{Myr}$ pre-MS isochrones of
  BCAH98 $\alpha=1.9$ (red, continuous), BCAH98 $\alpha=1.0$ (red, dashed),
  SDF00 (blue), DAM97 (green) and DCJ08 (cyan) are overlaid, adopting a distance modulus
  $dm=5.63$ and a reddening $E(B-V)=0.04$. The isochrones have been
  transformed into the observable plane using bolometric corrections
  derived by folding the atmospheric model flux distributions (see
  Section~\ref{atmospheric_models}) through the calculated INT-WFC
  system responses. The black line in each panel represents
  the spline fit (by eye) to the Pleiades single-star sequence. The
  open circles represent the photometric
  data, with the associated uncertainties in colour and magnitude
  shown as the bars.
  Top left: $K_{\rm{s}}, g_{_{\rm{WFC}}}-K_{\rm{s}}$ CMD. Top right:
  $g_{_{\rm{WFC}}}, (g-i)_{_{\rm{WFC}}}$ CMD. Bottom left:
  $r_{_{\rm{WFC}}}, (i-Z)_{_{\rm{WFC}}}$ CMD. Bottom right:
  $K_{\rm{s}}, J-K_{\rm{s}}$ CMD.}
\label{fig:pleiades_untuned_iso}
\end{figure*}

\subsection{Discussion}
\label{discussion_pleiades}

Fig.~\ref{fig:pleiades_untuned_iso} shows four CMDs of the Pleiades
with the pre-MS isochrones of BCAH98, SDF00, DAM97 and DCJ08
overlaid.  
The break between MS and pre-MS stars in the Pleiades occurs at $(g-i)_{_{\rm WFC}} \simeq 2$, and
is marked by a paucity of stars in all CMDs.
It is clear that the isochrones all follow the Pleiades MS reasonably well, though as we shall see
in Section~\ref{empirical_bolometric_corrections} this is in part is due to compensating errors in
each individual band.
None of the models, however, trace the observed pre-MS locus, with each predicting colours
that are too blue for a given stellar mass.
This issue has already been highlighted by \cite{Stauffer07} (see their Fig.~13), though from their 
data one could only be certain there was a problem in the $V$-band,
whereas our data show that all optical bands are affected.

This is part of a broader picture where it is known that MS stars are similarly affected, at least 
in the $V$-band (see
Section~\ref{optical_nir_mass-luminosity_relations}).
It is well known that there are sources of opacity missing from the
atmospheric models for cool stars ($T_{\rm{eff}} \lesssim 3700\,
\rm{K}$) in the optical regime of the spectrum
\citep{Leggett96,Baraffe98,Alvarez98,Baraffe02}. Although
the BT-Settl models are computed using more complete
line lists than previous generations of \textsc{phoenix}
models, there is
still a need for additional data to provide fully comprehensive line
lists, especially for the molecular opacities (H$_{2}$O, TiO,
CH$_{4}$, etc.; see \citealp{Partridge97} and
\citealp{Schwenke98}).

If we are to be quantitative about the missing opacity we need to measure the missing 
flux individually in each bandpass, rather than as a function of colour.  
The most straightforward way of achieving this is to identify a photometric bandpass 
where the missing sources of opacity are minimal, and create colours with respect to
that band.
This is equivalent to using the luminosity in that band to define the temperature.
To test the temperature scale requires masses (from which the models predict 
temperatures) from binaries, data which are simply not available for the Pleiades.
Instead, as the opacity problem affects MS stars as well as pre-MS stars (in the same
$T_{\rm{eff}}$ range), we will use
MS binaries to identify a band we can use. 

\section{Comparing the models and the data -- Main-Sequence Binaries}
\label{optical_nir_mass-luminosity_relations}

Although there are many MS binaries with well determined masses and distances 
we could use to test the models, the outstanding problem is the small sub-sample
of these which have individual magnitudes in many colours.
Astrometric binaries are often separated in the infrared (IR), and so lack individual optical
colours, and eclipsing binaries often have lightcurves in only a small number of colours.
To overcome this problem we will model the system magnitude, which as we shall
discuss later still gives significant insight as a function of stellar temperature.

\subsection{The sample}
\label{eclipsing_spectroscopic_binaries}

To create our sample, data for low-mass binaries from \cite{Delfosse00} 
were supplemented with data for higher-mass
binaries from \cite{Andersen87}; \cite{Clausen08}; \cite{Clausen09};
\cite{Clement97a}; \cite{Clement97b}; \cite{Lacy97}; \cite{Lacy05};
\cite{Lopez05}; \cite{Popper86}; \cite{Popper94}; \cite{Popper97};
\cite{Torres97}, and \cite{Torres09}.
We imposed an upper limit to the mass uncertainty on both
the primary and secondary components of
10~per~cent, and in addition limited ourselves to non-contact
binaries in which the effects of tidal forces and mass transfer may
bias the results. 
We also limited the sample to binaries with associated uncertainties in 
the parallax of $\leq 11$~per~cent;
thus our conclusions are unaffected by the Lutz-Kelker bias \citep{Lutz73}.
The sample span the mass range $0.1
\leq \msun \leq 1.4$ for the individual components, with the
higher mass limit chosen as the BCAH98 models only extend
to such masses. For the low-mass systems in \cite{Delfosse00}, the
apparent magnitudes were from the homogeneous dataset of
\cite{Leggett92}. A number of higher mass systems only had a Str{\"o}mgren $b-y$
colour and so these were transformed into $V-I_{\rm{c}}$ colours
according to the relation of \cite{Bessell79} which is accurate to
within $0.01\, \rm{mag}$ for MS stars. Near-IR measurements were taken from the 2MASS
point source catalogue \citep{Cutri03}. Parallax measurements
were taken from the Hipparcos catalogue \citep{Perryman97}, the
Yale General Catalogue of Trigonometric Stellar Parallaxes
\citep*{VanAltena95} and \cite{Segransan00}. 

\begin{figure}
\centering
\includegraphics[width=\columnwidth]{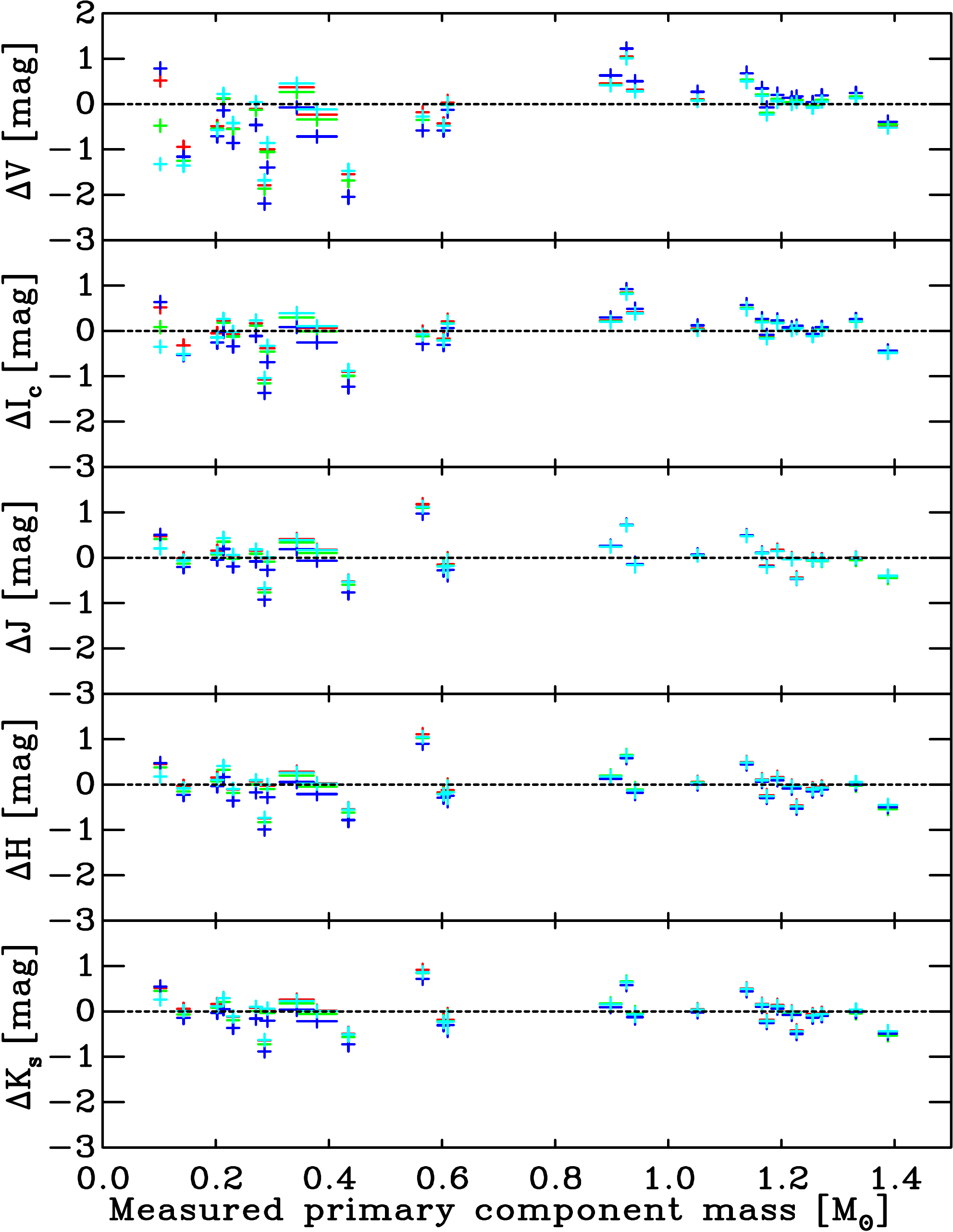}
\caption{The
  difference between the observed and theoretically predicted system
  absolute magnitudes (where $\Delta$ implies calculated -- observed)
  for the 28 binary systems as a function of the measured primary
  component stellar mass. Bolometric luminosities have been calculated
  using $2\, \rm{Gyr}$
  interior models of BCAH98 $\alpha=1.9$ (red), SDF00 (blue), DAM97
  (green) and DCJ08 (cyan). The atmospheric model flux distribution
  were folded through the revised $UBVRI$ responses of
  \protect\cite{Bessell12} and the $JHK_{\rm{s}}$ responses of
  \protect\cite{Cohen03} to create bolometric corrections which were
  used to derive the predicted absolute magnitudes.
  Upper panel: $\Delta V$. Upper middle panel:
  $\Delta I_{\rm{c}}$. Middle panel: $\Delta J$. Lower middle panel: $\Delta H$. Lower panel:
  $\Delta K_{\rm{s}}$.}
\label{fig:delta_vijk_mass}
\end{figure}

\subsection{The models}
\label{models}

We adopted $2\, \rm{Gyr}$
isochrones with which to compare observations against theory as we
expect these binaries, which are located in the field, to be old
systems. Note that at such ages, the mass scale is insensitive to changes
in age. For each binary component, the bolometric luminosity
$(L_{\rm{bol}})$, defined by the stellar interior models, was converted into an absolute magnitude using
bolometric corrections calculated using Eqn.~\ref{bc_final} by folding
the atmospheric models through the revised
$UBVRI$ responses of \cite{Bessell12}, and the $JHK_{\rm{s}}$
responses of \cite*{Cohen03}. For the reference Vega spectrum we used
the CALSPEC 
alpha\_lyr\_stis\_005\footnote[4]{\url{http://www.stsci.edu/hst/observatory/cdbs/calspec.html}}
spectrum with $V=0.03$ and all colours equal to zero. For the 2MASS
responses we used the bandpass specific reference spectra and
zero-point offsets of
\cite{Cohen03} (see their Table~2 and Section~5 respectively). The absolute magnitude was then converted
into a flux, and the fluxes for a given binary pair were then summed
and converted back into an absolute system magnitude. Where possible,
we have included the effects of interstellar reddening by adopting
literature values. These values range from $E(B-V)=0.0-0.11$. The
atmospheric models were reddened accordingly and the bolometric
corrections calculated as discussed in
Section~\ref{model_parameters}.

\subsection{Discussion}
\label{discussion_binaries}

Fig.~\ref{fig:delta_vijk_mass} shows the residuals (calculated $-$
observed) in luminosity relation in different bandpasses. 
As the binary mass ratio $q \rightarrow 1$ both
components have the same mass, whereas when $q
\rightarrow 0$ the primary component is likely to dominate the flux
output of the system. Thus we plot the difference in absolute
magnitude as a function of the measured primary mass. The range of $q$
as defined by our sample is $q=m_{2}/m_{1}=0.33-1.0$, where for 65~per~cent
of these the ratio $q > 0.9$.
Regardless of the choice of interior model, there are significant systematic
deviations (especially in the optical regime) for masses below $\simeq 0.4-0.5\,
\msun$. Fig.~\ref{fig:delta_vijk_mass} shows that the differences
between the observed and calculated system absolute magnitude
decreases as one moves to redder bandpasses, from approximately $2.5\,
\rm{mag}$ in the $V$-band to less than $1\, \rm{mag}$ in the
$K_{\rm{s}}$-band. In the near-IR, and in particular the $K_{\rm{s}}$-band, there is far less spread for a
given stellar mass (see also \citealp{Martin00} and
\citealp{Delfosse00}), with the interior models predicting similar
luminosities. This further highlights the fact that there is a problem
with the optical colours predicted by the atmospheric models for
masses below $\simeq 0.4-0.5\, \msun$ ($T_{\rm{eff}} \lesssim 3700\,
\rm{K}$).

It has been noted in the literature
that short-period, close-in binary system components have radii which
are inflated of order 10~per~cent with respect to evolutionary models
of low-mass MS stars (e.g. \citealp{Kraus11}). To test whether we
observe this effect in the photometry of these systems, we plotted
Fig.~\ref{fig:delta_vijk_mass} as a function of physical
separation. We found no evidence for the models to either under- or
overpredict the absolute system magnitude for short-period binaries
with smaller physical separations.

What is clear, however, is that of all the bands available, the $K_{\rm{s}}$-band magnitude
is closest to that predicted by the models for a given temperature, both in terms
of mean magnitude and spread.
This is unsurprising, since  the $K_{\rm{s}}$-band is less affected than either the 
$J$- or $H$-band by the missing sources of opacity that arise from incomplete H$_{2}$O
line lists (see the discussion on the completeness of the BT2 H$_{2}$O
line list as a function of wavelength by \citealp{Barber06}).
In what follows, therefore, we will use the $K_{\rm{s}}$-band magnitude to determine the
temperatures of Pleiades members.

\section{Quantifying the discrepancy}
\label{empirical_bolometric_corrections}

Our approach for quantifying the effects of the missing opacity is to determine
the temperature of a given Pleiad by comparing its $K_{\rm{s}}$-band magnitude 
with that predicted by a given model.
We can then compare the predicted luminosity in any other band with that of
the member in question to calculate the missing flux.
In practice it is better to work with a sequence than individual members and so
we defined the single-star Pleiades sequence in the INT-WFC and 2MASS colours by
fitting a spline (by eye) to the observed sequence in various
CMDs. 
We used only stars with uncertainties in both colour and magnitude
of less than $0.1\, \rm{mag}$. 
This spline lies slightly above the lower envelope of the
sequence to account for photometric uncertainties and takes into
consideration that the equal-mass binary sequence lies $\simeq 0.75\,
\rm{mag}$ above the single-star sequence (with very few systems of
higher multiplicity). 
Table~\ref{tab:pleiades_emp_sequence} provides the
single-star sequence in the $(griZ)_{_{\rm{WFC}}}JHK_{\rm{s}}$ bandpasses.

Starting, for instance, with the $K_{\rm{s}}$,
$(g_{_{\rm{WFC}}}-K_{\rm{s}})$ Pleiades CMD, we compared the
theoretical $130\, \rm{Myr}$ isochrone to the observed
spline and thus calculated the difference between the colours at a
given $T_{\rm{eff}}$ (provided by the model isochrone). Assuming that
the problem, in this case, lies in the $g_{_{\rm{WFC}}}$-band and not $K_{\rm s}$, this then
defines the required correction ($\Delta$BC; defined as
$\rm{BC}_{\rm{obs}} - \rm{BC}_{\rm{calc}}$) for the
$g_{_{\rm{WFC}}}$-band at a specific $T_{\rm{eff}}$. This process was
then repeated over the $T_{\rm{eff}}$ range of the theoretical
isochrone or the colour range specified by the observed spline, whichever was
more restrictive, thus providing $\Delta$BC as a function of
$T_{\rm{eff}}$ for the $g_{_{\rm{WFC}}}$-band. Repeating this for all
four pre-MS evolutionary models (including both the BCAH98
computations), we calculated the
model dependent $\Delta$BCs as a function of $T_{\rm{eff}}$ for the
$(griZ)_{_{\rm{WFC}}}JH$ bandpasses. 

\begin{figure*}
\centering
\includegraphics[width=0.9\textwidth]{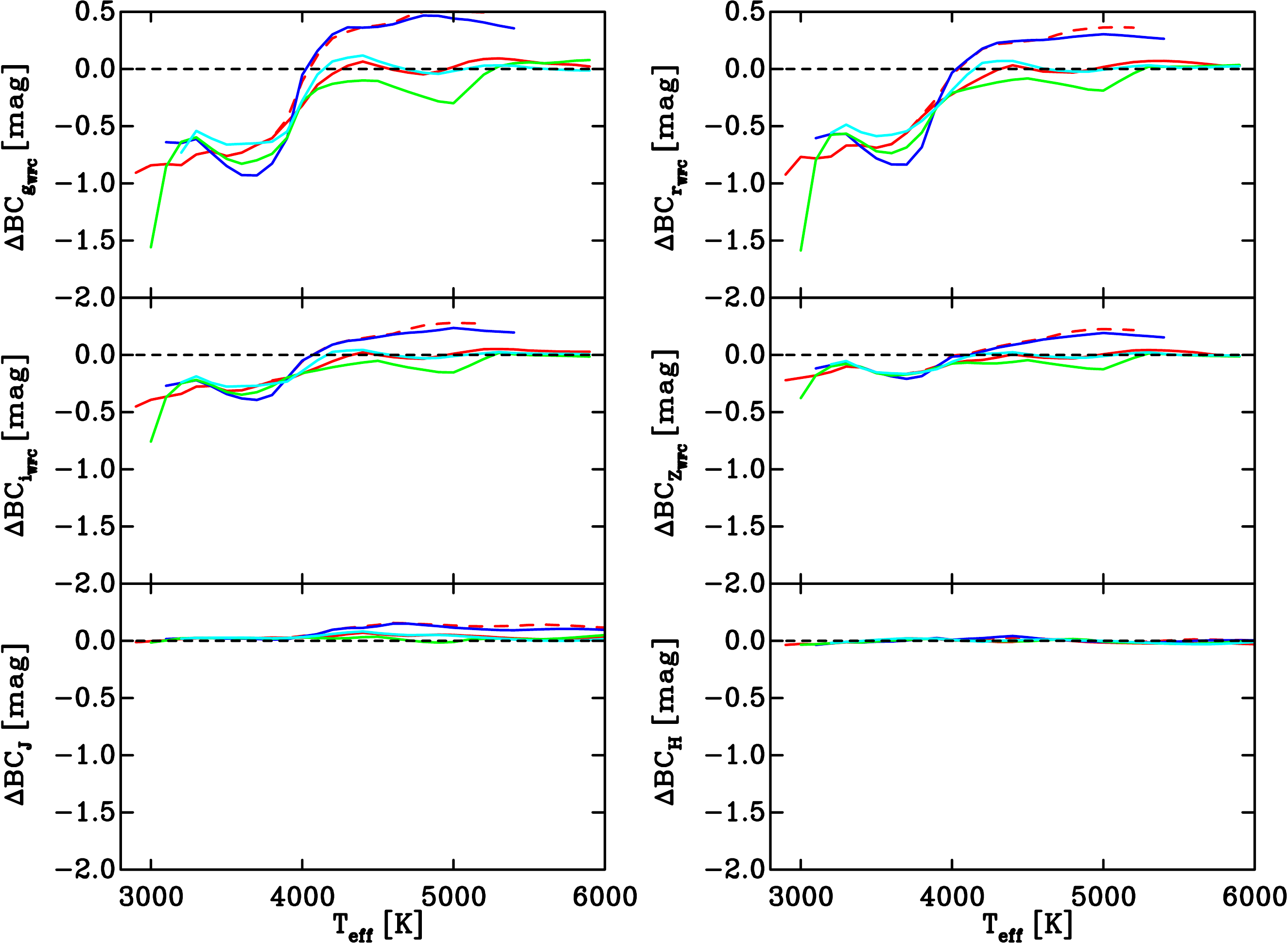}
\caption{Model dependent corrections ($\Delta$BC) calculated as a function
  of $T_{\rm{eff}}$ for the optical $(griZ)_{_{\rm{WFC}}}$ and near-IR $JH$
  bandpasses. The corrections were calculated as the difference
  between the theoretically predicted colour and the observed colour
  of the Pleiades sequence at a given $T_{\rm{eff}}$. Corrections have
  been calculated for the following models: BCAH98 $\alpha=1.9$ (red,
  continuous), BCAH98
  $\alpha=1.0$ (red, dashed), SDF00 (blue), DAM97 (green) and DCJ08
  (cyan). For $T_{\rm{eff}} \lesssim 4000\, \rm{K}$ all the models
  fail to match the data in the optical regime, with the magnitude of the
  mismatch decreasing with increasing wavelength.}
\label{fig:delta_bc_grizjh}
\end{figure*}

Our final results are shown in Fig.~\ref{fig:delta_bc_grizjh}.
In the Pleiades, stars cooler than $T_{\rm{eff}} \simeq 4000\, \rm{K}$ are on the
pre-MS, and all models fail to match the data in the optical in this regime.
The discrepancy is large, $0.75\, \rm{mag}$ in $g_{_{\rm{WFC}}}$, but
decreases with increasing wavelength. This is a clear improvement over
the \textsc{phoenix}/GAIA models, for which we find a
discrepancy of around $1\, \rm{mag}$.

We can test what effect the overestimation in flux of the atmospheric
models has on the derived ages for $T_{\rm{eff}} < 4000\, \rm{K}$ in
the following way. If this discrepancy is
independent of age (and therefore log$\,g$), the  
roughly logarithmic age spacing between the isochrones in CMD space
means that the resulting error in age is best represented as a
fractional rather than an absolute difference. We begin by
taking the BCAH98 $\alpha=1.9$ isochrones and adding our $\Delta$BCs derived above
to the theoretical bolometric corrections and colour-$T_{\rm{eff}}$
relations at a given age. We then
find the closest matching uncorrected BCAH98 $\alpha=1.9$ isochrone (in
CMD space). Comparing these, we find that for
ages between 3 and $10\, \rm{Myr}$, the difference in age is
approximately a factor 3 in $g, g-i$ and 2 in $r, r-z$.
 
The message here is very clear, one should use the reddest waveband possible to 
determine an age.
Unfortunately by the time one reaches the $JHK_{\rm{s}}$ bandpasses, the pre-MS isochrones are
vertical with a colour $J-K_{\rm{s}} \simeq 0.85$ for $2500 <
T_{\rm{eff}} < 4000\, \rm{K}$. Although individual stars descend
with age, the resulting sequences are almost degenerate with age, and for
young clusters ($< 10\, \rm{Myr}$), observations are further
complicated by the presence of discs. Thus the
$JHK_{\rm{s}}$ IR is unlikely to yield reliable young cluster ages.
Masses do not suffer from such a degeneracy and therefore the
$H$-band may be a reliable mass indicator when $K_{\rm{s}}$ is
affected by circumstellar material.

Perhaps more surprising is the fact that in the MS regime, hotter than 
$T_{\rm{eff}} \simeq 4000\, \rm{K}$, only two models fit the data in the optical.
The reason the BCAH98 $\alpha=1.0$ and SDF00 models tend to
overpredict the luminosity at a given $T_{\rm{eff}}$ is
that neither is tuned to the Sun, as explained in
Section~\ref{pre-ms_models}. Once the BCAH98 models have a tuned
mixing length paramater ($\alpha=1.9$) based on matching observed solar values they
are a good match to the MS data. A possible reason why the DAM97 models
fail to fit the MS may be because of their treatment of convection.

\begin{table}
\centering
\caption{The Pleiades single-star sequence in the INT-WFC and 2MASS bandpasses. The full table is
  available online, a sample is shown here as a representation of its content.}
\begin{tabular}{c c c c c c c}
\hline \hline
$g_{_{\rm{WFC}}}$&$r_{_{\rm{WFC}}}$&$i_{_{\rm{WFC}}}$&$Z_{_{\rm{WFC}}}$&$J$&$H$&$K_{\rm{s}}$\\
\hline
10.807&10.377&10.270&10.251&9.370&9.082&8.987\\
10.972&10.505&10.389&10.364&9.484&9.181&9.078\\
11.133&10.645&10.516&10.481&9.594&9.275&9.166\\
11.292&10.783&10.640&10.596&9.699&9.364&9.251\\
11.445&10.914&10.758&10.704&9.796&9.445&9.330\\
11.593&11.040&10.871&10.809&9.887&9.520&9.405\\
11.735&11.161&10.979&10.908&9.970&9.588&9.473\\
11.870&11.273&11.079&11.000&10.043&9.646&9.534\\
11.998&11.378&11.172&11.086&10.110&9.699&9.589\\
12.120&11.478&11.261&11.168&10.169&9.745&9.637\\
\hline \hline
\end{tabular}
\label{tab:pleiades_emp_sequence}
\end{table}

\section{Conclusions}
\label{conclusions}

We have carried out a precise test of a set of pre-MS models by
examining how well they match the Pleiades sequence in optical and near-IR CMDs.
The stages we have gone through to achieve this are as follows.

\begin{enumerate}
\item{We have demonstrated that traditional photometric
    calibration, using observations of standard stars to
    transform the data into a standard photometric system, should not be used for
    studies of pre-MS stars. 
    Differences in log$\,g$ and $T_{\rm{eff}}$
    between a MS and pre-MS star of the same colour can, for instance,
    result in errors of order $0.1-0.15\, \rm{mag}$ in
    both the $g$-band magnitude and $g-i$ colour for red stars. Hence
    it is crucial that precise photometric studies (especially of
    pre-MS objects) be carried out in the natural photometric
    system of the observations.}\\
\item{Therefore we have calculated system responses for the natural INT-WFC photometric 
    system. Our observations show these are a good model of the 
    photometric system.}\\
\item{We have demonstrated that for all optical colours no pre-MS
    model follows the observed Pleiades sequence for temperatures
    cooler than $ 4000\, \rm{K}$.
    The models overestimate
    the flux by a factor two at $0.5\, \mu$m, with the difference decreasing with
    increasing wavelength. We believe there is little observable effect at
    $2.2\, \mu$m, as the $K_{\rm{s}}$-band magnitude of MS
    binaries matches the models at this wavelength.}\\
\item{These differences in bolometric correction correspond to underestimating the 
    ages of pre-MS stars younger than $10\, \rm{Myr}$ by factors of two to three, depending on 
    the choice of colours used.
    Thus the errors in pre-MS isochrone models can explain the discrepancy between MS
    and pre-MS ages found by \cite{Naylor09}, and implies the ages determined from the
    MS are more likely to be correct.}\\
\item{We have provided both our system responses and our Pleiades sequence in such
    a way that it can be used as a benchmark for any new evolutionary
    models. These are given in
    Tables~\ref{tab:int-wfc_system_responses} and
    \ref{tab:pleiades_emp_sequence} respectively and will also be made
    available from the Cluster Collaboration
    homepage\footnote[5]{\url{http://www.astro.ex.ac.uk/people/timn/Catalogues/}}
    and the CDS archive.}
\end{enumerate}

\section*{Acknowledgements}

CPMB is funded by a UK Science and Technology Facilities Council
(STFC) studentship. SPL is supported by an RCUK fellowship.
This research has made use of data obtained at the
\textit{Isaac Newton} Telescope which is operated on the island of La
Palma by the \textit{Isaac Newton} Group (ING) in the Spanish
Observatorio del Roque de los
Muchachos of the Institutio de Astrofisica de Canarias.
This research has made use of archival data products from the
Two-Micron All-Sky Survey (2MASS), which is a joint project of the
University of Massachusetts and the Infrared Processing and Analysis
Center, funded by the National Aeronautics and Space Administration (NASA)
and the National Science Foundation.

\bibliographystyle{mn3e}
\bibliography{references}

\appendix

\section{The calculated INT-WFC system responses}
\label{creating_int/wfc_system_responses}

\begin{figure}
\centering
\includegraphics[width=\columnwidth]{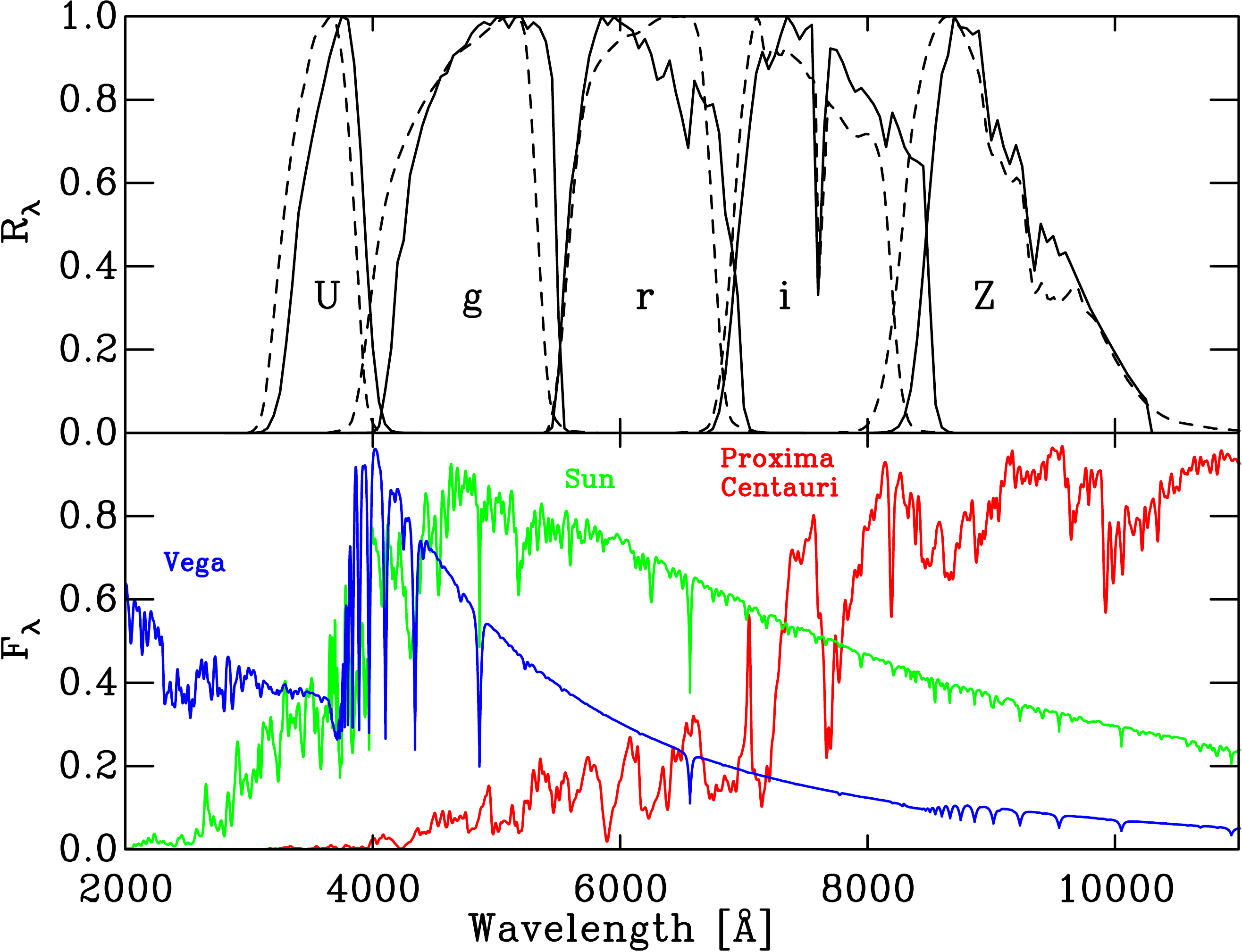}
\caption{The normalised system response functions for the Sloan Digital
  Sky Survey (dashed) and the Wide-Field Camera on the \textit{Isaac
    Newton} Telescope (INT-WFC; bold). The INT-WFC system responses
  are those calculated in
  Appendix~\ref{creating_int/wfc_system_responses} and include the
  effects of the telescope optics and atmospheric absorption. Plotted in
  the bottom panel are normalised model spectra of Vega (A0V), the Sun (G2V) and
  Proxima Centauri (M6V).}
\label{fig:filters}
\end{figure}

Although the SDSS survey system responses are
well constrained and available on the SDSS
webpages\footnote[6]{\url{http://www.sdss.org/dr7/instruments/imager/index.html#filters}}
\citep{Doi10},
the responses on the \textit{Isaac Newton} Group (ING) webpages for the WFC
only combine the filter throughput and CCD quantum
efficiency. To model
the system responses of the INT-WFC we included
the cumulative effects of the transmission of the Earth's atmosphere, the
reflectivity of the telescope mirror, the transmission of the prime
focus corrector
optics, the quantum efficiency of the detector, and the filter transmission.
To calculate the
atmospheric transmission we used the model for the La Palma atmosphere
derived by
\cite{King85}, for an airmass typical for our observations of
1.4. This model varies smoothly as a function of wavelength and
does not include the molecular absorption features. These atmospheric
absorption bands (primarily due to H$_{2}$O, CO$_{2}$ and O$_{3}$)
were estimated using the spectrum of an F8 star observed using the
Faint Object Spectrograph on the \textit{William Herschel}
Telescope. At low resolution the continuum of an F8 star is relatively
smooth and thus the atmospheric bands can be identified by fitting a
low order polynomial to the continuum and subtracting this from the
spectrum \citep{Shahbaz96}. The continuum is modelled between $4700 -
9800\, \rm{\AA}$. At prime focus the INT optical path
involves a single reflection from the aluminium coated primary mirror
and this was modelled using the aluminium reflectivity spectrum of
\cite{Allen63}. A three-element prime focus corrector is
used which is coated to minimise reflection and thus improve
efficiency and reduce ghosts. The first two elements are non-interchangeable and have a
broadband single layer MgF$_{2}$ coating. The third element, although
interchangeable, typically adds an additional
single layer of MgF$_{2}$ coating. These single layer coatings produce total
reflectivities smaller than 2~per~cent in the wavelength region $3500 -
7400\, \rm{\AA}$. Beyond this range, the
reflectivity from the prime focus corrector increases to $\simeq
3$~per~cent at $8000\, \rm{\AA}$.
The filter and detector responses were from the ING
webpages\footnote[7]{\url{http://www.ing.iac.es/Astronomy/instruments/wfc/}}.
The detector is comprised of four EEV42-80 CCDs. The detector response on
the ING webpages only extends to $3400\, \rm{\AA}$ in the blue,
however the blue edge of the RGO $U_{_{\rm{WFC}}}$ filter is at approximately $3000\,
\rm{\AA}$ and so the detector response was extended blueward
using data for an almost identical EEV44-82 CCD to
a wavelength of $3200\, \rm{\AA}$ and further extrapolated to
$3000\, \rm{\AA}$ \citep{Cavadore00}.
The RGO $Z_{_{\rm{WFC}}}$ filter response data becomes noisy at
wavelengths greater than $9000\, \rm{\AA}$ and so this response was
extended redward using calculated 4-mm Schott RG850 data to $\simeq 10\, 300\,
\rm{\AA}$ where the detector response approaches zero. Any negative
values for the filter responses from the ING webpages were
removed. Note also that the RGO $U_{_{\rm{WFC}}}$ filter response used here does
not include the recently discovered red leak near $7050\, \rm{\AA}$
\citep{Verbeek12}. The calculated INT-WFC system responses are shown
in the top panel of Fig.~\ref{fig:filters}, with the individual
bandpass throughputs given in
Table~\ref{tab:int-wfc_system_responses}.

\begin{table*}
\caption{The normalised calculated INT-WFC system responses.}
\begin{tabular}{c c c c c c c c c c}
\hline \hline
$\lambda$&$U_{_{\rm{WFC}}}$&$\lambda$&$g_{_{\rm{WFC}}}$&$\lambda$&$r_{_{\rm{WFC}}}$&$\lambda$&$i_{_{\rm{WFC}}}$&$\lambda$&$Z_{_{\rm{WFC}}}$\\
\hline
3050&0.000&4000&0.000&5400&0.000&6700&0.000&8050&0.000\\
3100&0.002&4050&0.015&5450&0.013&6750&0.010&8100&0.001\\
3150&0.011&4100&0.099&5500&0.142&6800&0.050&8150&0.002\\
3200&0.042&4150&0.203&5550&0.384&6850&0.142&8200&0.005\\
3250&0.097&4200&0.409&5600&0.586&6900&0.285&8250&0.015\\
3300&0.216&4250&0.463&5650&0.684&6950&0.462&8300&0.041\\
3350&0.356&4300&0.618&5700&0.812&7000&0.605&8350&0.104\\
3400&0.528&4350&0.676&5750&0.904&7050&0.745&8400&0.223\\
3450&0.614&4400&0.736&5800&0.957&7100&0.863&8450&0.391\\
3500&0.694&4450&0.781&5850&1.000&7150&0.915&8500&0.570\\
3550&0.767&4500&0.813&5900&0.980&7200&0.873&8550&0.736\\
3600&0.832&4550&0.855&5950&0.998&7250&0.906&8600&0.861\\
3650&0.902&4600&0.864&6000&0.986&7300&0.956&8650&0.906\\
3700&0.954&4650&0.905&6050&0.976&7350&1.000&8700&1.000\\
3750&1.000&4700&0.918&6100&0.966&7400&0.978&8750&0.974\\
3800&0.992&4750&0.930&6150&0.923&7450&0.945&8800&0.971\\
3850&0.885&4800&0.959&6200&0.946&7500&0.968&8850&0.956\\
3900&0.684&4850&0.976&6250&0.909&7550&0.980&8900&0.966\\
3950&0.425&4900&0.986&6300&0.848&7600&0.331&8950&0.829\\
4000&0.206&4950&0.979&6350&0.856&7650&0.736&9000&0.702\\
4050&0.074&5000&0.996&6400&0.894&7700&0.923&9050&0.751\\
4100&0.021&5050&0.998&6450&0.817&7750&0.919&9100&0.687\\
4150&0.005&5100&0.972&6500&0.758&7800&0.889&9150&0.645\\
4200&0.001&5150&0.997&6550&0.684&7850&0.848&9200&0.691\\
4250&0.000&5200&1.000&6600&0.845&7900&0.819&9250&0.640\\
&&5250&0.970&6650&0.809&7950&0.828&9300&0.493\\
&&5300&0.985&6700&0.784&8000&0.809&9350&0.390\\
&&5350&0.972&6750&0.789&8050&0.789&9400&0.502\\
&&5400&0.939&6800&0.721&8100&0.759&9450&0.458\\
&&5450&0.851&6850&0.526&8150&0.686&9500&0.473\\
&&5500&0.285&6900&0.455&8200&0.769&9550&0.426\\
&&5550&0.011&6950&0.329&8250&0.731&9600&0.433\\
&&5600&0.000&7000&0.061&8300&0.686&9650&0.400\\
&&&&7050&0.008&8350&0.660&9700&0.366\\
&&&&7100&0.003&8400&0.652&9750&0.333\\
&&&&7150&0.001&8450&0.641&9800&0.299\\
&&&&7200&0.000&8500&0.341&9850&0.272\\
&&&&&&8550&0.068&9900&0.245\\
&&&&&&8600&0.013&9950&0.219\\
&&&&&&8650&0.003&$10\, 000$&0.192\\
&&&&&&8700&0.000&$10\, 050$&0.167\\
&&&&&&&&$10\, 100$&0.141\\
&&&&&&&&$10\, 150$&0.119\\
&&&&&&&&$10\, 200$&0.097\\
&&&&&&&&$10\, 250$&0.080\\
&&&&&&&&$10\, 300$&0.000\\
\hline \hline
\end{tabular}
\label{tab:int-wfc_system_responses}
\end{table*}

\section{Calculating the theoretical transformations}
\label{relative_differences_synthetic_photometry}

To investigate what effect variations in the system responses between
the standard SDSS and natural INT-WFC photometric systems would have
on the calibration of our photometric data, we calculated the
magnitude and colour differences between the two photometric systems.
We adopted the standard SDSS responses, the calculated INT-WFC
responses and folded the atmospheric models through both. For this comparison we were
interested only in the relative differences in the derived colours and
magnitudes, and so our conclusions are not be affected by the use of
atmospheric models.

Following the formalism of \cite{Girardi02}, a synthetic apparent
magnitude $m_{_{R_{\lambda}}}$ in a given filter with
response function $R_{\lambda}$ is calculated as,

\begin{equation}
m_{_{R_{\lambda}}} = -2.5\, \mathrm{log} \left(\frac{\int_{\lambda}
    \lambda f_{\lambda} R_{\lambda} d\lambda}{\int_{\lambda} \lambda
    f^\circ_{\lambda} R_{\lambda} d\lambda}\right) + m^\circ_{_{R_{\lambda}}},
\label{app_mag}
\end{equation}

\noindent where $f_{\lambda}$ is the stellar flux as observed at Earth
and $f^\circ_{\lambda}$ denotes a reference spectrum with a
known apparent magnitude $m^\circ_{_{R_{\lambda}}}$. The response function
$R_{\lambda}$ is the product of the component responses
described in Appendix~\ref{creating_int/wfc_system_responses}. Note that the integrands in
Eqn.~\ref{app_mag} imply photon counting across the filter which is
appropriate for detectors such as CCDs.

Stellar atmospheric models only contain information about the flux
at the stellar surface, $F_{\lambda}$. The transformation from
$f_{\lambda}$ to $F_{\lambda}$ requires knowledge of the
stellar radius, its distance and an
extinction curve $A_{\lambda}$ to account for interstellar
extinction. To incorporate these and remove the radial dependence, it
is more convenient to deal in terms of bolometric corrections, which
can be formulated as,

\begin{eqnarray}
\label{bc_final}
  BC_{_{R_{\lambda}}} & = & M_{\rm{bol},\odot} -
  2.5\, \mathrm{log}\left(\frac{4\pi(10\mathrm{pc})^{2}F_{\rm{bol}}}{L_{\odot}}\right) \\
  & + &~2.5\, \mathrm{log}\left(\frac{\int_{\lambda}\lambda
      F_{\lambda}10^{-0.4 A_{\lambda}}R_{\lambda}\, d\lambda}{\int_{\lambda}\lambda
    f^\circ_{\lambda}R_{\lambda}\, d\lambda}\right) -
   m^\circ_{_{R_{\lambda}}}, \nonumber
\end{eqnarray}

\noindent where $F_{\rm{bol}} = \sigma T^{4}_{\rm{eff}}$ is
the total flux emergent at the stellar surface and all other symbols
retain their original definitions. For the solar values we used
$M_{\rm{bol}, \odot} = 4.74$ and $L_{\odot} = 3.855 \times 10^{33}\,
\rm{erg\, s^{-1}}$ \citep*{Bessell98}.

The SDSS photometric system is based on a monochromatic
AB magnitude system (see \citealp{Oke83,Fukugita96}) which by
definition means that a reference spectrum of constant flux density
per unit frequency $f^\circ_{\nu}$ will have AB magnitudes
$m^\circ_{\nu} = 0$ at all frequencies $\nu$.
By converting this reference spectrum in Eqn.~\ref{bc_final} to energy
per unit wavelength according to
$f_{\lambda} = c/\lambda^{2}\, f_{\nu}$, this
definition can then be extended to any photometric system with the result that
$m^\circ_{\lambda} = 0$ are the reference magnitudes. Thus the
magnitude and colour differences for a given flux distribution between
the INT-WFC and SDSS photometric systems were then simply calculated
as the difference in the derived bolometric corrections.

We required bolometric corrections for the MS and pre-MS
regimes. To model the changing $T_{\rm{eff}}$ and
log$\,g$ values along the MS (from log$\,g$ $\simeq$ 4
for early B-type stars to $\simeq$ 5.5 for late M-type) we
used the calculated ZAMS of
SDF00 for stellar masses in the range $0.1 - 7\,
\msun$. For the pre-MS values we adopted the SDF00 interior models for
ages of 1 and $10\, \rm{Myr}$, calculating $T_{\rm{eff}}$ and log$\,g$ over the same mass range.
The reason for choosing the SDF00 interior models
over other pre-MS stellar interior models was simply because they span a
greater range of stellar masses and so more thoroughly map the
MS. Due to the
degree of model dependency in pre-MS evolutionary models, we repeated
the calculation of the theoretical transformations using the ZAMS
$T_{\rm{eff}}$, log$\,g$ values from the BCAH98, DAM97 and DCJ08 models and found that
all four sets of evolutionary models agree to within $0.02\,
\rm{mag}$ over the range of stellar masses.

Fig.~\ref{fig:delta_wfc_sdss_red} shows the calculated magnitude and
colour differences for reddened and unreddened objects (for
illustrative purposes we adopted a nominal reddening of $E(B-V)=1$). The simple formalism
adopted in the derivation of the bolometric corrections allowed us to
include the effects of interstellar reddening by applying a given
extinction curve $A_{\lambda}$ and calculating
the bolometric correction as shown in Eqn.~\ref{bc_final}. We adopted
the \cite{Cardelli89} extinction curve
of $R_{V} = A_{V}/E(B-V)$ and thus derived the extinction in a
given bandpass at a specific $E(B-V)$. We adopted $R_{V}=3.2$ as this
value best reproduced the colour dependent reddening vector of
\cite{Bessell98} upon folding the \textsc{atlas9} ``no-overshoot''
atmospheric models \citep{Kurucz92} through the $UBVRI$ responses
of \cite{Bessell90}. Magnitude and colour differences were
calculated for each $T_{\rm{eff}}$, log$\,g$ combination in both the
MS and pre-MS regimes via linear interpolation.

\label{lastpage}

\end{document}